\documentclass[11pt]{article}

\usepackage{amstext,amsgen,latexsym,amsmath}
\usepackage{amstext,amssymb,amsfonts,latexsym}
\usepackage{theorem}
\usepackage{pifont}
\usepackage{times}

\usepackage{graphics,epsfig}

 \newcommand{\hs}[1]{\hspace*{ #1 mm}}

\theoremstyle{plain}
\theorembodyfont{\rmfamily}
 \newtheorem{theorem}{Theorem}[section]
 \newtheorem{lemma}[theorem]{Lemma}
 \newtheorem{proposition}[theorem]{Proposition}

 {\theorembodyfont{\rmfamily}
\newtheorem{definition}[theorem]{Definition}}
 {\theorembodyfont{\rmfamily} }
 {\theorembodyfont{\rmfamily} }
 {\theorembodyfont{\rmfamily} }

 \newtheorem{claim}{Claim}

 \newenvironment{proof}{\par \noindent
            {\bf Proof. \hs{2}}}{\hfill$\Box$ \vspace*{3mm}}

 \newcommand{\nat}{\mathbb{N}}
 \newcommand{\integer}{\mathbb{Z}}

 \newcommand{\QSCDFF}{QSCD$_{\it ff}$}
 \newcommand{\QSCDC}{QSCD$_{\it cyc}$}
 \newcommand{\DIST}{DIST}
 \newcommand{\uap}{\mathrm{\bf UAP}}
 \newcommand{\spp}{\mathrm{\bf SPP}}
 \newcommand{\np}{\mathrm{{\bf NP}}}
 \newcommand{\conp}{\mathrm{co}\mbox{-}\mathrm{\bf NP}}
 
 \newcommand{\coam}{\mathrm{co}\mbox{-}\mathrm{\bf AM}}

 \newcommand{\bra}[1]{\langle #1 |}
 \newcommand{\ket}[1]{| #1 \rangle}
 
 \newcommand{\ketbra}[2]{| #1 \rangle\langle #2 |}

 \newcommand{\Ker}{\mathrm{Ker}}

 \def\magicwand{\itemsep=0pt\parskip=0pt\topskip=0pt}
 \newcommand{\ignore}[1]{}

\begin{document}
%%%%%%%%%%%%%%%%%%%%
\sloppy
\title{Computational Indistinguishability between Quantum States\\ and Its Cryptographic Application\footnote{The preliminary version~\cite{KKNY05} appeared 
in the Proceedings of EUROCRYPT 2005, Lecture Notes in Computer Science, Vol.3494, pp.268--284, Aahus, Denmark, May 22--26, 2005.}}

\author{
Akinori Kawachi$^{1}$$\qquad$
Takeshi Koshiba$^{2}$$\qquad$
Harumichi Nishimura$^{3}$$\qquad$
Tomoyuki Yamakami$^{4}$\\
\\
$^{1}$ Department of Mathematical and Computing Sciences,
Tokyo Institute of Technology\\ 
$^{2}$ Division of Mathematics, Electronics and Informatics\\
Graduate School of Science and Engineering, Saitama University\\
$^{3}$ Department of Mathematics and Information Sciences\\
Graduate School of Science, Osaka Prefecture University\\
$^{4}$ ERATO-SORST Quantum Computation and Information Project\\
Japan Science and Technology Agency\\
}
\date{}
%\begin{singlespace}
\maketitle
%\end{singlespace}

%%%%%%%%%%%%%%%%%%%%%%%%%%%%%%%%%%%%%%%%%%%%%%%%%%%%%%%%%%%%%%%%%%%%
\begin{abstract}
We introduce a computational problem of distinguishing between two specific quantum states as a new 
cryptographic problem to design a quantum cryptographic scheme that is ``secure'' against any polynomial-time  
quantum adversary. Our 
problem, \QSCDFF, is to distinguish 
between two types of random coset states with a hidden permutation over
the symmetric group of finite degree. This naturally generalizes the 
commonly-used distinction problem between two probability distributions
 in computational cryptography. As our major contribution, we show 
that \QSCDFF\ has three properties of cryptographic interest: 
($i$) \QSCDFF\ has a trapdoor; 
($ii$) the average-case hardness of \QSCDFF\ coincides 
with its worst-case hardness; and ($iii$)  \QSCDFF\ is computationally at least 
as hard as the graph automorphism problem in the worst case. 
These cryptographic properties enable us to construct a quantum public-key cryptosystem, which is likely to withstand any chosen plaintext attack of a polynomial-time 
quantum adversary. We further discuss a generalization of \QSCDFF , called \QSCDC , and introduce a multi-bit encryption scheme that relies 
on similar cryptographic properties of \QSCDC .
\end{abstract}

%%%%%%%%%%%%%%%%%%%%%%%%%%%%%%%%%%%%%%%%%%%%%%%%%%%%%%%%%%%%%%%%%%%%
\section{Introduction}\label{intro}

In 1976, Diffie and Hellman \cite{DH76} first used a computationally intractable 
problem to design a key exchange protocol. Computational cryptography 
has since then become an important field of extensive study. 
A number of practical cryptographic 
systems (e.g., public-key cryptosystems (PKCs), bit commitment schemes (BCSs), 
pseudorandom generators, and digital signature schemes) have been proposed  
under popular intractability assumptions, such as the hardness of the 
integer factorization problem (IFP) and the discrete logarithm problem (DLP), for which no efficient classical %(i.e., deterministic or probabilistic) 
algorithm has been found.
Using the power of quantum computation, however, we can efficiently solve 
various number-theoretic problems, including IFP (and thus, the quadratic residuosity problem) 
\cite{Sho97},  DLP (and also the Diffie-Hellman problem) \cite{BL95,Kit95,Sho97}, 
and the principal ideal problem \cite{Hal07} (see also \cite{Des02,Sch09}). 
This indicates that a {\em quantum adversary} (i.e., an adversary who  operates a quantum computer) 
can easily break any cryptosystems whose security proofs rely on the computational hardness of 
those problems.

In order to deal with such a powerful quantum adversary, a new area of cryptography, so-called {\em quantum cryptography}, has emerged in the past quarter century. In 1984, Bennett and 
Brassard~\cite{BB84} first proposed a {\em quantum key distribution scheme\/}, in which a party can securely send a secret key to another party through a quantum communication channel. Its unconditional security was later 
proven by Mayers \cite{May01} (and more sophisticated proofs were 
given  by, e.g., Shor and Preskill~\cite{SP00} and Renner~\cite{Ren05}). 
Against our early hope, quantum mechanics cannot make 
all cryptographic schemes information-theoretically secure since, 
for instance, as Mayers \cite{May97} and Lo and Chau \cite{LC97} 
independently demonstrated, no quantum BCS can be both 
unconditionally concealing and binding. 
Therefore, ``computational'' approaches are still important 
and also viable in quantum cryptography. Along this line of study, a number of quantum cryptographic properties have been discussed from complexity-theoretic viewpoints  
\cite{AC02,CDMS04,CLS01,DFS04,DMS00,OTU00}.

In fact, a quantum computer is capable of breaking the RSA cryptosystem and 
many other well-known classical cryptosystems. It is therefore imperative to discover  computationally-hard problems from which we can construct 
a quantum cryptosystem that is 
secure against any polynomial-time quantum adversary.  
For instance, the subset sum (knapsack) problem and the shortest vector problem are used as bases of knapsack-based cryptosystems \cite{IN96,OTU00} as well as lattice-based cryptosystems \cite{AD97,Reg04b,Reg09}.
Since we do not know whether these problems  
withstand any attack by quantum adversaries, we need to continue searching for better intractable problems that can guard their associated quantum cryptosystems against any computationally-bounded quantum adversary. 

This paper naturally generalizes a notion of the 
computational indistinguishability between two probability distributions 
\cite{BM84,GM84,Yao82} to that between 
two {\em quantum states}. In particular, 
we present a distinction problem, called \QSCDFF\ (quantum state computational distinction with 
fully flipped permutations), between specific ensembles of quantum 
states. It turns out that \QSCDFF\ enjoys useful cryptographic properties as a building block of a secure quantum cryptosystem. Henceforth, $\nat$ denotes the set of all non-negative integers.

\begin{definition}\label{q-ind}
The {\em advantage\/} of a polynomial-time quantum algorithm ${\cal A}$ that 
distinguishes between two ensembles $\{\rho_0(l)\}_{l\in\nat}$
and $\{\rho_1(l)\}_{l\in\nat}$ of quantum states is the function $\delta_{\cal A}(l)$ defined 
as: 
\[
 \delta_{\cal A}(l) = \left| \Pr_{\cal A}[{\cal A}(\rho_0(l))=1] - \Pr_{\cal A}[{\cal A}(\rho_1(l))=1] \right|
\]
for two $l$-qubit quantum states $\rho_0(l)$ and $\rho_1(l)$,
where the subscript ${\cal A}$ of the probability means that any output of 
${\cal A}$ is 
determined by measuring the final state of ${\cal A}$ in the standard 
computational basis.
We say that two ensembles $\{\rho_0(l)\}_{l\in\nat}$ and 
$\{\rho_1(l)\}_{l\in\nat}$ are {\em computationally indistinguishable\/} if the advantage $\delta_{\cal A}(l)$ is negligible for any polynomial-time 
quantum algorithm ${\cal A}$; namely, for any polynomial $p$, any 
polynomial-time quantum algorithm ${\cal A}$, and any sufficiently large number $l$, it holds that $\delta_{\cal A}(l)<1/p(l)$.
The distinction problem between $\{\rho_0(l)\}_{l\in\nat}$ 
and $\{\rho_1(l)\}_{l\in\nat}$ is 
said to be {\em solvable with non-negligible advantage\/} if 
these ensembles are not computationally indistinguishable; that is, 
there exist a polynomial-time quantum algorithm ${\cal A}$ and a 
polynomial $p$ such that
\[
 \left| \Pr_{\cal A}[{\cal A}(\rho_0(l))=1] - \Pr_{\cal A}[{\cal A}(\rho_1(l))=1] \right| > \frac{1}{p(l)}
\]
for infinitely many numbers $l$.
\end{definition}

%Watrous \cite{Wat06} recently gave another definition of computational 
%indistinguishability between two quantum states
%based on a non-uniform model, i.e., a quantum adversary with quantum advice,
%for computational zero-knowledge protocols against quantum attacks.
%On the contrary, our definition is based on the uniform computational model for 
%other cryptographic purposes such as public-key cryptosystems.

Let $N=\{n\in\nat :\ \text{$n$ is even and $n/2$ is odd}\} = 
\{n\in\nat: n\equiv 2\ (\bmod\ 4)\}$.
The problem \QSCDFF\ 
asks whether an adversary can distinguish between 
two sequences of identical copies of $\rho_\pi^+(n)$ and of 
$\rho_\pi^-(n)$,
where $n$ is a length parameter in $N$ and $\pi$ is unknown to the adversary. For each $n\in N$, let $S_n$ denote 
the {\em symmetric group} of degree $n$ and let 
${\cal K}_n = \{\pi\in S_n : \pi^2=id\,\,\mbox{and}\,\,\forall i\in\{1,...,n\}[\pi(i)\neq i] \}$, where $id$ stands for the identity permutation. 
We say a permutation is odd if it can be expressed 
by an odd number of transpositions, and even otherwise.
Denote by ${\rm sgn}$ the {\em sign function} of permutations,  
defined as ${\rm sgn}(\pi)=0$ if $\pi$ is even and ${\rm sgn}(\pi)=0$ if $\pi$
is odd. Notice that,  for each $n\in N$,   
${\rm sgn}(\pi)=1$ for every $\pi\in{\cal K}_n$ 
(i.e., $\pi\in{\cal K}_n$ is an odd permutation)  
since $\pi$ consists of $n/2$ disjoint transpositions; in other words, 
it holds that $\pi=(i_1\ i_2)(i_3 i_4)\cdots(i_{n-1}\ i_n)$ 
for $n$ distinct numbers $i_1,\ldots,i_n$ in $\{1,...,n\}$.
This simple fact will be used for certain properties of \QSCDFF.

\begin{definition}\label{def:QSCDff}
For each $\pi\in {\cal K}_n$, let $\rho_\pi^+(n)$ and $\rho_\pi^-(n)$ be two quantum states defined by
\[
 \rho_\pi^+(n) = \frac{1}{2n!}\sum_{\sigma\in S_n}(\ket{\sigma}+\ket{\sigma\pi})(\bra{\sigma}+\bra{\sigma\pi})\mbox{~and~}
 \rho_\pi^-(n) = \frac{1}{2n!}\sum_{\sigma\in S_n}(\ket{\sigma}-\ket{\sigma\pi})(\bra{\sigma}-\bra{\sigma\pi}).
\]
The problem \QSCDFF\ is the distinction problem between two quantum states $\rho_\pi^+(n)^{\otimes k(n)}$ and 
$\rho_\pi^-(n)^{\otimes k(n)}$ for each parameter $n$ in $N$, where $k$ is a polynomial. For each fixed polynomial $k$, we use the succinct notation $k$-\QSCDFF\ instead. 
\end{definition}

To simplify our notation, we often drop the parameter $n$ whenever it is clear from the context. For instance, we write $\rho_{\pi}^{+\otimes k}$ instead of $\rho_{\pi}^{+}(n)^{\otimes k(n)}$. 
More generally, $k$-\QSCDFF\ can be defined for any integer-valued 
function $k$.
Note that Definition \ref{def:QSCDff} uses the parameter $n$ to express  
the ``length'' of the quantum states instead of the parameter $l$ 
of Definition~\ref{q-ind}. Speaking of polynomial-time indistinguishability, however, there is essentially no difference  between $n$ and $l$ because $\rho_\pi^+$ and 
$\rho_\pi^-$ can be expressed by $O(n\log{n})$ qubits and $k(n)$ 
is a polynomial in $n$. 
In this paper, the parameter $n$ serves as a unit of the computational complexity of our target 
problem and it is often referred to as the {\em security parameter\/} in a cryptographic context. 

%%%%%%%%% 
\subsection{Our Contributions}
This paper presents three properties of \QSCDFF\ and their direct 
implications 
toward building a secure quantum cryptographic scheme. 
These properties are summarized as follows. 
({\it i})~\QSCDFF\ has a {\em trapdoor}; namely, we can efficiently distinguish between $\rho_\pi^+$ and $\rho_\pi^-$ 
if $\pi\in{\cal K}_n$ is known. 
 ({\it ii})~The average-case hardness of \QSCDFF\ over a randomly 
chosen permutation 
$\pi\in {\cal K}_n$ coincides with its worst-case hardness. 
({\it iii})~\QSCDFF\ is computationally at least as hard in the worst case as the {\em graph automorphism problem} (GA), where GA is the graph-theoretical problem defined as: 
\begin{quote}
{\sc Graph Automorphism Problem (GA):}\\
{\sf input:} an undirected graph $G = (V,E)$, where $V$ is a set of nodes and $E$ is a set of edges;\\
{\sf output:} YES if $G$ has a non-trivial automorphism, and NO 
otherwise.
\end{quote}
Since there is no known efficient algorithmic solution for GA, the third property suggests that \QSCDFF\ should be difficult to solve in polynomial time.
We are also able to show, without any 
assumption, that no  
time-unbounded quantum algorithm can solve $o(n\log{n})$-\QSCDFF .
Making use of the aforementioned three cryptographic properties, we can design a computationally-secure  
quantum PKC whose security relies on the worst-case hardness of GA. The following subsection will discuss in depth numerous advantages of using \QSCDFF\ as a basis of secure quantum cryptosystems.

As a further generalization of \QSCDFF , we present another distinction problem \QSCDC , which satisfies the following cryptographic 
properties: ({\it i}) it has a trapdoor
and ({\it ii}) its average-case hardness coincides with the worst-case hardness. This new problem becomes a basis for another public-key cryptosystem that can encrypt messages longer than those encrypted by 
the encryption scheme based on \QSCDFF .

%%%%%%%%%
\subsection{Comparison between Our Work and Previous Work}\label{sec:comparison}
In a large volume of the past literature, 
computational-complexity aspects of quantum states have been spotlighted in connection to quantum cryptography. 
In the context of 
quantum zero-knowledge proofs, for instance, the notion of statistical distinguishability between 
two quantum states was investigated by Watrous \cite{Wat02} and also by 
Kobayashi \cite{Kob03}. 
They proved that certain problems of statistical  
distinction between two quantum states 
are promise-complete for quantum 
zero-knowledge proof systems. Concerning the computational complexity 
of quantum-state generation, Aharonov and Ta-Shma \cite{AT07} studied its direct connection to quantum adiabatic 
computing as well as statistical zero-knowledge proofs. 
In a similar vein, our distinction problem \QSCDFF\ is also rooted in computational complexity 
theory. 

In the remaining of this subsection, we briefly discuss various advantages of using \QSCDFF\ as a basis of quantum cryptosystems by comparing 
it with the underlying problems of existing cryptosystems. 

\paragraph{Average-Case Hardness versus Worst-Case Hardness.}
For any given problem, its efficient solvability on average does not, in general, guarantee that the problem should be solved efficiently 
even in the worst case. 
Consider the following property of cryptographic problems:  
the average-case hardness of the problem is ``equivalent'' to its worst-case hardness under a certain type of polynomial-time reduction. 
Since the worst-case hardness of the problem is much more desirable, 
this average-case/worst-case property certainly increases our confidence in the security of 
the cryptographic scheme. 
Unfortunately, few cryptographic problems are known to enjoy this property.

In the literature, there are two major categories of worst-case/average-case reductions. 
The first category involves a {\em strong reduction}, which transforms an arbitrary instance of length $n$ to a random instance of the same length  $n$ or rather length polynomial in 
$n$.  With this strong reduction, Ajtai \cite{Ajt96} found a remarkable connection between average-case hardness 
and worst-case hardness of certain variants of the so-called 
shortest vector problem (SVP). 
He gave an efficient reduction from a problem of approximating
the shortest vector of a given $n$-dimensional lattice in the worst case 
to another problem of approximating the shortest vector 
of a random lattice 
within a larger approximation factor. 
Later, Micciancio and Regev 
\cite{MR07} established a much better average-case/worst-case 
connection with respect to the approximation of SVP. 

Unlike the first one, the second category is represented by a {\em weak  reduction} of Tompa and Woll \cite{TW87}, where the reduction is randomized only over a certain portion of all the instances. 
A typical example is DLP, which can be randomly reduced to itself by a reduction that maps instances not to all instances of the same length but rather to all instances of the same underlying group.
Concerning DLP, it is not known whether an efficient reduction exists 
from DLP with the worst-case prime to DLP 
with a random prime. By Shor's algorithm \cite{Sho97}, we can  
efficiently solve DLP as well as the inverting problem of the 
RSA function, which have worst-case/average-case reductions 
of the second category. The graph isomorphism 
problem (GI) and the aforementioned GA---well-known graph-theoretical problems---also satisfy weak worst-case/average-case reductions 
\cite{TW87} although there is no known cryptosystem whose security relies on their hardness. See \cite{BT06} 
and references therein for more information on worst-case/average-case reductions. 

In this paper, we show that \QSCDFF\ has a worst-case/average-case 
reduction of the {\em first category}.  Unlike the reduction of DLP, our reduction depends only on the size of each instance. In fact, our distinction problem \QSCDFF\ is the {\em first} cryptographic problem 
having a worst-case/average-case 
reduction of the first category; namely, 
the worst case of the problem can be reduced to the average case of the {\em same} problem. 
Our reduction is similar in flavor to the reductions used for 
the aforementioned lattice problems. 
In the case of the approximation of SVP, however, 
an approximation problem of SVP can be reduced randomly only to 
{\em another} approximation problem with a worse parameter.
Note that, on a quantum computer, no efficient solution is currently known for \QSCDFF\ . 

\paragraph{Computational Hardness of Underlying Computational Problems.}
The hidden subgroup problem (HSP) has played a central role in various discussions on the strengths and limitations of 
quantum computation.
The aforementioned IFP and DLP can be reduced to special cases of HSP 
on Abelian groups (AHSP). Kitaev \cite{Kit95} showed how to solve  AHSP efficiently; in particular, he gave a polynomial-time algorithm 
that performs the quantum 
Fourier transformation over Abelian groups, 
which is a generalization of the quantum Fourier transformation 
used in, e.g., Shor's algorithm \cite{Sho97}. 
To solve HSP on non-Abelian groups, a simple application of currently known techniques may not be sufficient despite of the existence of an efficient quantum algorithm for AHSP. 
Notice that, over certain specific non-Abelian groups, HSP was already solved in \cite{BCD05b,EH00,GSVV04,HRT03,Kup05,MRRS04,Reg04a}. 
Another important variant of HSP is HSP on the dihedral groups (DHSP). 
Regev \cite{Reg04a}
demonstrated a quantum reduction from the 
unique shortest vector problem (uSVP)
to a slightly different variant of DHSP, where uSVP can serve as  
a basis of lattice-based PKCs defined in \cite{AD97,Reg04b}.
A subexponential-time quantum algorithm for DHSP was found by Kuperberg \cite{Kup05}. Although these results do not immediately give a desired  
subexponential-time quantum algorithm for uSVP, 
it could eventually lead us to design the desired algorithm.

Our problem \QSCDFF\ is closely related to
another variant: HSP on the {\em symmetric groups} (SHSP), which appears to be much more difficult to solve than the aforementioned 
variants of HSP do. Note
that no known subexponential-time quantum algorithm exists for SHSP. 
Recently, Hallgren, Russell, and Ta-Shma~\cite{HRT03} introduced a 
distinction problem, similar to \QSCDFF , 
between certain two quantum states to discuss 
the computational intractability of SHSP by a ``natural'' extension of 
Shor's algorithm \cite{Sho97}.
In this paper, we refer to their distinction problem as \DIST. 
An efficient solution to \DIST\ gives rise to an efficient quantum 
algorithm for a certain special case of SHSP.
To solve \DIST, as they showed, we require 
exponentially many trials of the so-called {\em weak Fourier sampling} 
that works on a single copy of the quantum states. In other words, 
exponentially many copies are needed in total as far as 
the weak Fourier sampling is used. 

This result was improved by  
Grigni, Schulman, Vazirani, and Vazirani~\cite{GSVV04}, who proved 
that exponentially many copies are necessary even if we use 
a powerful method, known as {\em strong Fourier sampling},  
along with a random choice of the bases of the representations
of the symmetric group $S_n$. 
Concerning the computational hardness
of SHSP, Kempe and Shalev \cite{KS05} further expanded 
the results of \cite{GSVV04,HRT03} with quantum Fourier sampling methods. 
Moore, Russell, and Schulman~\cite{MRS08}, on the contrary, 
demonstrated that, regardless of
the method (such as the above quantum Fourier sampling methods), any 
time-unbounded quantum algorithm working on a single copy needs $\exp(\Omega(n))$ 
trials to solve \DIST. Even for the case of two copies, 
Moore and Russell \cite{MR05} argued that   
any time-unbounded quantum algorithm that simultaneously works over two
copies requires $\exp (\Omega(\sqrt{n}/\log{n}))$ trials at best. 
Their results were further improved by 
Hallgren, Moore, R{\"o}tteler, Russell, and Sen~\cite{HMRRS06}, who 
proved that no time-unbounded quantum algorithm solves \DIST\ 
even if it simultaneously works over $o(n\log{n})$ copies. 
In this paper, we show that 
the distinction problem \DIST\ is, in fact, 
polynomial-time reducible to \QSCDFF . 
This immediately implies, from the above results,  
that no quantum algorithm 
solves \QSCDFF\ using $o(n\log{n})$ copies.
% a new part in 2nd submission

Even by supplying sufficiently many copies to an algorithm, 
there is no known subexponential-time quantum algorithm that solves 
\QSCDFF, and thus
%On the other hand, if we ignore the time complexity, their distinction problem can be actually solved with $O(n\log{n})$ copies by the result of Ettinger et al.~\cite{EHK04}. Even if their result is applicable to \QSCDFF , we currently have no efficient, even subexponential-time, quantum algorithms for \QSCDFF .
finding such an algorithm seems a 
daunting task. This situation indicates that our problem, \QSCDFF, is 
much more suitable than, for example, uSVP for an underlying intractable problem to build a secure cryptosystem. 
There is a similarity with the classical case of 
DLP over different groups; namely, DLP over $\integer_p^*$ (where $p$ is a prime) is 
classically computable in subexponential time whereas no known classical subexponential-time  
algorithm exists for DLP over certain groups in elliptic curve cryptography. {}From this reason, it is generally believed that DLP over such groups is more reliable than DLP over $\integer_p^*$.

%Our problem \QSCDFF\ is closely related to
%a much harder problem: HSP on the symmetric groups (SHSP). No subexponential-time quantum algorithm is known for SHSP.
%A distinction problem, similar to \QSCDFF , defined in terms of SHSP 
%was introduced by Hallgren, Russell and Ta-Shma \cite{HRT03}, who showed that 
%no quantum algorithms solve their problem in polynomial time using the standard method\footnote{The algorithms that run an essential part of Shor's algorithm \cite{Sho97} are simply called {\em  standard methods}.}.
%The hardness result of Hallgren et al.~was also strengthened by \cite{GSVV04,KS05,HRS05,MR05}.
%Here, we show that their problem is polynomial-time reducible to \QSCDFF . This immediately implies that any standard algorithm that solves \QSCDFF\ also 
%requires exponential time. The hardness result of Hallgren et al.  was recently strengthened by Grigni et al.~\cite{GSVV04} and Kempe and Shalev \cite{KS05}.  
%Finding even a subexponential algorithm for \QSCDFF\ seems a daunting task. On the contrary, this suggests that our problem \QSCDFF\ is more reliable than, e.g., uSVP.
%This situation is similar to the case of DLP over different groups on  
%classical computation. DLP over $\integer_p^*$ ($p$ is a prime) is 
%classically solved in subexponential time whereas there is no known classical subexponential-time  
%algorithm for DLP over certain groups used in elliptic curve cryptography. It is believed that DLP over such groups is more reliable than DLP over $\integer_p^*$.

We prove that the computational complexity of \QSCDFF\ is lower-bounded by that of GA. Well-known upper bounds of GA include $\np\cap\coam$ \cite{GS89,Sch88}, $\spp$ \cite{AK06}, and $\uap$ \cite{CGRS04}; however, 
GA is not known to sit in 
$\np\cap\conp$. Notice that, since most cryptographic problems fall in $\np\cap\conp$, very few 
cryptographic systems are lower-bounded by the worst-case hardness of problems outside of $\np\cap\conp$.

\paragraph{Quantum Computational Cryptography.}
Apart from PKCs, quantum key distribution gives a foundation
to symmetric-key cryptology; for instance, the quantum key distribution scheme in \cite{BB84} achieves unconditionally
secure sharing of secret keys in symmetric-key cryptosystems (SKCs) through 
an authenticated classical communication channel and an insecure quantum 
communication channel.
Undoubtedly, both SKCs and PKCs have their own advantages and disadvantages.
Compared with SKCs,
PKCs require fewer secret keys in a large-scale network; however, they often need certain intractability  
assumptions for their security proofs and are typically vulnerable to, e.g.,  
the man-in-the-middle attack. As an immediate application of \QSCDFF , 
we  propose a new 
computational quantum PKC whose security relies on the computational 
hardness of \QSCDFF . 

Of many existing PKCs, few make their security proofs solely rely on the {\em worst-case} hardness of their underlying problems, 
such as lattice-based 
PKCs (see, e.g., \cite{Reg09}).
A quantum adversary is a powerful foe who can easily break many PKCs whose underlying problems are number-theoretic, because these problems can be  efficiently solved on a quantum computer. Based on a certain subset of the knapsack problem, Okamoto, Tanaka, and Uchiyama~\cite{OTU00} 
proposed a quantum PKC which withstands certain well-known quantum 
attacks.
Our proposed quantum PKC also seems to fend off 
a polynomial-time quantum adversary since 
we can reduce the problem GA to \QSCDFF , where GA
is not known to be solved efficiently on a quantum computer.

\subsection{Later Work}
After the publication of the preliminary version \cite{KKNY05} 
of this paper, the notion of quantum-state indistinguishability and its associated quantum encryption schemes have been further studied.  
Here are some of the recent results related to the topics of this paper. 
Hayashi, Kawachi, and Kobayashi~\cite{HKK08} showed that 
\QSCDC\ satisfies the indistinguishability property against 
time-unbounded quantum algorithms in such a way that \QSCDFF\ does.
In information-theoretical settings, Nikolopoulos~\cite{Nik08} and Nikolopoulos and Ioannou~\cite{NI09} proposed
new quantum encryption schemes.
Kawachi and Portmann~\cite{KP08} proved that, with respect to the 
ratio of message length and key size,  any quantum encryption scheme 
has no advantage over a classical one-time pad scheme if we impose 
certain information-theoretically strong security requirement 
on the quantum encryption scheme.

%%%%%%%%%%%%%%%%%%%%%%%%%%%%%%%%%%%%%%%%%%%%%%%%%%%%%%%%%%%%%%%%%%%%
\section{Cryptographic Properties of \QSCDFF}\label{sec:property}

Through this section, we will show that \QSCDFF\ enjoys the following 
three cryptographically useful  
properties: ({\it i}) a trapdoor, ({\it ii})
the equivalence between average-case hardness and worst-case hardness under polynomial-time reductions, and ({\it iii}) 
a reduction from two computationally-hard problems to \QSCDFF.
These properties will help us to construct a quantum PKC in Section~\ref{sec:application}. We assume, throughout this paper, the reader's familiarity with the basics of quantum computation \cite{NC00} and 
of finite group theory \cite{Rob95}.

All the cryptographic properties of \QSCDFF\ are consequences of the following characteristics of the set ${\cal K}_n$ of the hidden permutations. ({\it i})~Each permutation $\pi\in{\cal K}_n$ is of order 2. This provides the 
trapdoor of \QSCDFF . ({\it ii})~For any $\pi\in {\cal K}_n$, the 
conjugacy class $\{\tau^{-1}\pi\tau: \tau\in S_n \}$
of $\pi$ is equal to ${\cal K}_n$. This property enables us to prove the equivalence 
between the worst-case hardness and average-case hardness of \QSCDFF .
({\it iii})~The problem GA is (polynomial-time Turing) equivalent to its subproblem with 
the promise that any given graph has either 
a unique non-trivial automorphism in 
${\cal K}_n$ or none at all. This equivalence relation is used to give a 
complexity-theoretic lower bound of \QSCDFF ; that is, 
the average-case hardness of \QSCDFF\ is lower-bounded by 
the worst-case hardness of GA. 
To prove those properties, we introduce two new techniques: ({\it i}) a variant of the 
so-called {\em coset sampling method\/}, which is widely used in various  
extensions of Shor's well-known algorithm (see, e.g., \cite{Reg04a}) and ({\it ii}) a quantum version of the 
{\em hybrid argument}, which is a powerful tool for many security reductions used in computational  
cryptography.

Now, recall the two quantum states 
$
 \rho_\pi^+ = \frac{1}{2n!}\sum_{\sigma\in S_n} (\ket{\sigma}+\ket{\sigma\pi})(\bra{\sigma}+\bra{\sigma\pi})$ and $\rho_\pi^- = \frac{1}{2n!}\sum_{\sigma\in S_n} (\ket{\sigma}-\ket{\sigma\pi})(\bra{\sigma}-\bra{\sigma\pi})
$ 
for a permutation $\pi\in{\cal K}_n$. For convenience, let $\iota(n)$ (or simply $\iota$) denote the maximally mixed state  
$\frac{1}{n!}\sum_{\sigma\in S_n}\ketbra{\sigma}{\sigma}$ over $S_n$, which will appear later.

%%%%%%%%%%%%%%%%%%%%%%%%%%%%%%%%%%%
\subsection{A Trapdoor}

We start by proving that \QSCDFF\ has a {\em trapdoor}. To prove 
this claim, 
it suffices to present an efficient distinguishing 
algorithm between $\rho_\pi^+$ and $\rho_\pi^-$ with an extra knowledge 
of their hidden permutation~$\pi\in{\cal K}_n$. 

\begin{theorem}[Distinguishing Algorithm]\label{trapdoor}
There exists a polynomial-time quantum 
algorithm that, for any security parameter $n\in N$ and for any hidden permutation $\pi\in {\cal K}_n$, distinguishes between 
$\rho_\pi^+(n)$ and $\rho_\pi^-(n)$ using $\pi$ with probability $1$.
\end{theorem}

\begin{proof}
Fix $n\in N$ arbitrarily. Let $\chi$ be any given unknown quantum 
state, which is limited to either 
$\rho_\pi^+$ or $\rho_\pi^-$. 
The desired distinguishing algorithm for $\chi$ works as follows.
\begin{description}
\magicwand
\item[{\rm (D1)}] Prepare two quantum registers. The first register holds a control bit and the second register holds $\chi$. Apply the Hadamard transformation $H$ to the first register. 
The state of the system now becomes
\[
 H\ketbra{0}{0}H\otimes\chi.
\]
\item[{\rm (D2)}] Apply the Controlled-$\pi$ operator $C_{\pi}$ 
to the both registers, where the operator $C_{\pi}$ behaves as 
$
 C_{\pi}\ket{0}\ket{\sigma} = \ket{0}\ket{\sigma}
$
and 
$
 C_{\pi}\ket{1}\ket{\sigma} = \ket{1}\ket{\sigma\pi}
$
 for any given $\sigma\in S_n$.
Since $\pi^2 = id $ for every $\pi\in {\cal K}_n$,
the state of the entire system can be expressed as 
\[
 \frac{1}{n!}\sum_{\sigma\in S_n}\ketbra{\psi_{\pi,\sigma}^+}{\psi_{\pi,\sigma}^+}
\quad\text{if $\chi = \rho_\pi^+$,}\quad\text{and}\quad
 \frac{1}{n!}\sum_{\sigma\in S_n}\ketbra{\psi_{\pi,\sigma}^-}{\psi_{\pi,\sigma}^-}
\quad\text{if $\chi = \rho_\pi^-$,}
\]
where $\ket{\psi_{\pi,\sigma}^{+}}$ and $\ket{\psi_{\pi,\sigma}^{-}}$ are defined as 
\begin{eqnarray*}
 \ket{\psi_{\pi,\sigma}^\pm} &=& C_\pi\left(\frac{1}{2}\ket{0}\left(\ket{\sigma}\pm\ket{\sigma\pi}\right)+\frac{1}{2}\ket{1}\left(\ket{\sigma}\pm\ket{\sigma\pi}\right)\right)\\
                             &=& \frac{1}{2} \ket{0}(\ket{\sigma} \pm \ket{\sigma\pi}) + \frac{1}{2}\ket{1}(\ket{\sigma\pi} \pm \ket{\sigma}).
\end{eqnarray*}
\item[{\rm (D3)}] Apply the Hadamard transformation again 
to the first register. Since $\chi$ is either $\rho_\pi^+$ or $\rho_\pi^-$, the state of the entire system becomes either 
\[
 (H\otimes I)\ket{\psi_{\pi,\sigma}^+} = \frac{1}{\sqrt{2}}\ket{0}\left(\ket{\sigma} + \ket{\sigma\pi} \right)
\quad\text{or}\quad
 (H\otimes I)\ket{\psi_{\pi,\sigma}^-} = \frac{1}{\sqrt{2}}\ket{1}\left(\ket{\sigma} - \ket{\sigma\pi} \right),
\]
respectively. Measure the first register in the computational basis. 
If the measured result is $0$, then output YES; otherwise, output NO.
\end{description}
It is clear that the above procedure gives the correct answer with probability $1$.
\end{proof}

%%%%%%%%%%%%%%%%
\subsection{A Reduction from Worst Case to Average Case}

We intend to reduce the worst-case hardness of \QSCDFF\ to 
its average-case hardness.
Such a reduction implies that \QSCDFF\ with a random permutation $\pi$ is at least as hard as \QSCDFF\ with the fixed permutation $\pi'$ of the highest complexity. Since the converse reduction is 
trivial, the average-case hardness of \QSCDFF\ is therefore   polynomial-time Turing equivalent 
to its worst-case hardness.

\begin{theorem}\label{wa-reduction}
Let $k$ be any polynomial and let ${\cal A}$ be 
a polynomial-time quantum algorithm 
that solves $k$-\QSCDFF\ with non-negligible advantage for a uniformly 
random $\pi\in{\cal K}_n$; namely, there exists a polynomial $p$ such that, 
for infinitely many security parameters $n$ in $N$,
\[
 \left| \Pr\limits_{\pi, {\cal A}}[{\cal A}(\rho_\pi^{+}(n)^{ \otimes k(n)}) = 1] 
      - \Pr\limits_{\pi, {\cal A}}[{\cal A}(\rho_\pi^{-}(n)^{ \otimes k(n)}) = 1] \right| > \frac{1}{p(n)},
\]
where $\pi$ is chosen uniformly at random from ${\cal K}_n$.
Then, there exists a polynomial-time quantum algorithm ${\cal B}$ that 
solves $k$-\QSCDFF\ with non-negligible advantage for any permutation $\pi\in{\cal K}_n$.
\end{theorem}
\begin{proof}
Fix an arbitrary parameter $n\in N$ that satisfies the assumption of the theorem. Assume that our input is either $\rho_\pi^{+}(n)^{ \otimes k(n)}$  or $\rho_\pi^{-}(n)^{ \otimes k(n)}$. For each $i\in\{1,2,...,k(n)\}$, let $\chi_i$ be the $i$th state of the given $k(n)$ states. 
Clearly, $\chi_i$ is either $\rho_\pi^+$ or $\rho_\pi^-$. 
{}From the given average-case algorithm ${\cal A}$, we build the desired worst-case algorithm ${\cal B}$ in the following way.
\begin{description}
\magicwand
\item[{\rm (R1)}] Choose a permutation $\tau\in S_n$ uniformly at random.
\item[{\rm (R2)}] Apply $\tau$ to each $\chi_i$, where $i\in\{1,...,k(n)\}$, from the right.
If $\chi_i=\rho_\pi^+$, then we obtain the quantum state
\begin{eqnarray*}
 \chi_i' &=& \frac{1}{2n!}\sum_{\sigma\in S_n}(\ket{\sigma\tau}+\ket{\sigma\tau\tau^{-1}\pi\tau})(\bra{\sigma\tau}+\bra{\sigma\tau\tau^{-1}\pi\tau})\\
       &=& \frac{1}{2n!}\sum_{\sigma'\in S_n}(\ket{\sigma'}+\ket{\sigma'\tau^{-1}\pi\tau})(\bra{\sigma'}+\bra{\sigma'\tau^{-1}\pi\tau}).
\end{eqnarray*}
When $\chi_i=\rho_\pi^-$, we instead obtain
$\displaystyle
 \chi_i' = \frac{1}{2n!}\sum_{\sigma'\in S_n}(\ket{\sigma'}-\ket{\sigma'\tau^{-1}\pi\tau})(\bra{\sigma'}-\bra{\sigma'\tau^{-1}\pi\tau})$.
\item[{\rm (R3)}] Invoke the average-case quantum algorithm ${\cal A}$ on the input 
$\bigotimes_{i=1}^{k} \chi_i'$.
\item[{\rm (R4)}] Output the outcome of ${\cal A}$.
\end{description}
Let $\pi\in {\cal K}_n$. Note that, for each $\tau\in S_n$,  $\tau^{-1}\pi\tau$ belongs to ${\cal K}_n$. 
Moreover, for every $\pi'\in{\cal K}_n$, there exists a 
$\tau\in S_n$ satisfying $\tau^{-1}\pi\tau = \pi'$, from which 
it follows that the conjugacy class 
$\{\tau^{-1}\pi\tau: \tau\in S_n\}$ of $\pi$ is equal to ${\cal K}_n$. 
As shown below, the 
number of all permutations $\tau\in S_n$ for which $\tau^{-1}\pi\tau=\pi'$ is independent of the choice of $\pi'\in{\cal K}_n$.
\begin{claim}
For any permutations $\pi,\pi',\pi'' \in{\cal K}_n$,  
$|\{\tau\in S_n: \tau^{-1}\pi\tau=\pi' \}|=|\{\tau\in S_n: \tau^{-1}\pi\tau=\pi'' \}|$.
\end{claim}
\begin{proof}
Define a map $\mu_\tau:{\cal K}_n\rightarrow {\cal K}_n$ as 
$\mu_\tau(\sigma) = \tau^{-1}\sigma\tau$ and a set 
${\cal T}_{\pi,\pi'} := \{\mu_\tau: \mu_\tau(\pi)=\pi'\}$. 
It is obvious that, by defining  
a group operation ``$\,\cdot\,$'' as  
$\mu_\tau\cdot\mu_{\tau'}(\cdot) = \mu_\tau(\mu_{\tau'}(\cdot))$, 
${\cal T}_{\pi,\pi}$ becomes a subgroup of ${\cal S}_n:=\{\mu_\tau: \tau\in S_n\}$. Therefore, 
${\cal S}_n$ has a coset decomposition with respect to its subgroup 
${\cal T}_{\pi,\pi}$ for any $\pi\in {\cal K}_n$ and 
each coset coincides with 
${\cal T}_{\pi,\pi'}$ for a certain $\pi'$. This shows  that 
$|{\cal T}_{\pi,\pi'}|=|{\cal T}_{\pi,\pi''}|$ for every pair $\pi',\pi''$.
Since $\mu_\tau$ and $\tau$ have a one-to-one correspondence, 
it follows that,  for every $\pi',\pi''$,  
$|\{\tau\in S_n: \tau^{-1}\pi\tau=\pi' \}|=|\{\tau\in S_n: \tau^{-1}\pi\tau=\pi'' \}|$.
\end{proof}

The above-mentioned properties imply that  
$\tau^{-1}\pi\tau$ is indeed uniformly distributed over ${\cal K}_n$. Therefore, by 
feeding the input $\bigotimes_{i=1}^{k}\chi_i'$ to the algorithm ${\cal A}$, we can achieve the desired non-negligible advantage of ${\cal A}$. This completes the proof. 
\end{proof}

%%%%%%%%%%%%%%%%%%%%%%%%%%%%%%%%%%%
\subsection{Computational Hardness}

The third property of \QSCDFF\ relates to the computational hardness of \QSCDFF . We want to present two claims that witness its relative hardness against GA. 
First, we prove 
that the computational complexity of  \QSCDFF\ is lower-bounded 
by that of GA by constructing an efficient reduction from GA to \QSCDFF. Second,
we briefly discuss relationships among  \QSCDFF , SHSP, and DIST, and we then prove that \QSCDFF\ cannot be solved from $o(n\log{n})$ copies of input instances.

Now, we prove the first claim concerning the reducibility between  
GA and  \QSCDFF . 
Our reduction from GA to \QSCDFF\ 
consists of two parts: a reduction from GA to a variant of GA, called 
UniqueGA$_{\it ff}$, and a reduction from UniqueGA$_{\it ff}$ to \QSCDFF .
To describe the desired reduction, we formally introduce UniqueGA$_{\it ff}$. 
Earlier, K{\"o}bler, Sch{\"o}ning, and Tor{\'a}n~\cite{KST93} introduced the following 
{\em unique graph automorphism problem} (UniqueGA).
\begin{quote}
{\sc Unique Graph Automorphism Problem} (UniqueGA):\\
{\sf input:} an undirected graph $G = (V,E)$, where $V$ is a set of nodes and $E$ is a set of edges; \\
{\sf promise:} $G$ has either a unique non-trivial automorphism or no non-trivial automorphism;\\
{\sf output:} YES if $G$ has the non-trivial automorphism, and NO \hbox  to 0pt{otherwise.}
\end{quote}
Note that this promise problem UniqueGA is called (1GA, GA) in \cite{KST93}.
The {\em unique graph automorphism with fully-flipped permutation} (UniqueGA$_{\it ff}$) is a slight modification of UniqueGA. 
Recall that $N=\{n'\in\nat:\ n'\equiv 2\ (\bmod 4)\}$. 
\begin{quote}
{\sc Unique Graph Automorphism with Fully-Flipped Permutation} (UniqueGA$_{\it ff}$):\\
{\sf input:} an undirected graph $G = (V,E)$, where $V$ is a set of nodes and $E$ is a set of edges;\\
{\sf promise:} the number $n=|V|$ of nodes is 
in $N$. Moreover,
$G$ has either a unique non-trivial automorphism $\pi\in {\cal K}_{n}$
or no non-trivial automorphism;\\ 
{\sf output:} YES if $G$ has the non-trivial automorphism, and NO otherwise.
\end{quote}
Note that every instance $G$ of UniqueGA$_{\it ff}$ 
is defined only when the 
number $n$ of nodes belongs to $N$.

Regarding UniqueGA$_{\it ff}$, we want to prove two helpful lemmas. The first lemma uses a variant of a so-called {\em coset sampling method}, which has 
been widely used in many generalizations of Shor's algorithm. Recall that $\iota(n) = \frac{1}{n!}\sum_{\sigma\in S_n}\ketbra{\sigma}{\sigma}$ for each $n\in N$. 
\begin{lemma}
\label{positive-coset-sampling}
There exists a polynomial-time quantum algorithm that, given an instance $G$ of UniqueGA$_{\it ff}$, 
generates a quantum state $\rho_\pi^+$ if $G$ is an ``YES" instance 
with its unique non-trivial automorphism $\pi$, or generates $\iota$ 
if $G$ is a ``NO" instance.
\end{lemma}
\begin{proof}
Let $n\in N$. Given an instance $G$ of UniqueGA$_{\it ff}$, we first prepare the quantum state 
$
 \frac{1}{\sqrt{n!}}\sum_{\sigma\in S_n}\ket{\sigma}\ket{\sigma(G)},
$
where $\sigma(G)$ is the graph resulting from 
relabeling its nodes according to each permutation $\sigma$.
By discarding the second register, we can obtain a quantum state $\chi$ in the first register. 
If $G$ is an ``YES" instance with the unique non-trivial 
automorphism $\pi$, then 
this state $\chi$ equals $\rho_\pi^+$ since 
$
\frac{1}{\sqrt{n!}}\sum_\sigma \ket{\sigma}\ket{\sigma(G)} = 
\frac{1}{\sqrt{n!}}\sum_{\sigma\in S_n/\langle\pi\rangle} 
(\ket{\sigma}+\ket{\sigma\pi})\ket{\sigma(G)}.
$
Otherwise, 
since $\sigma(G)\ne\sigma'(G)$ for any distinct 
$\sigma,\sigma'\in S_n$, $\chi$ equals $\iota = \frac{1}{n!}\sum_{\sigma\in S_n}\ketbra{\sigma}{\sigma}$.
\end{proof}

The second lemma requires a variant of the coset sampling method as a technical 
tool. The lemma in essence relies on the fact that the hidden 
$\pi\in{\cal K}_n$ is an odd permutation for each $n\in N$ since, 
as a special property of ${\cal K}_n$,  
$\pi$ can be expressed as a product of an odd number of transpositions. 

\begin{lemma}\label{negative-coset-sampling}
There exists a polynomial-time quantum algorithm that, given an instance 
$G$ of UniqueGA$_{\it ff}$, generates a quantum state $\rho_\pi^-$ if $G$ is 
an ``YES'' instance with the unique non-trivial automorphism $\pi$  
or generates $\iota$ if $G$ is a ``NO'' instance.
\end{lemma}

\begin{proof}
Let $n\in N$. Similar to the algorithm given in the proof 
of Lemma~\ref{positive-coset-sampling}, we start with the quantum state
$
 \frac{1}{\sqrt{n!}}\sum_{\sigma\in S_n}\ket{\sigma}\ket{\sigma(G)}
$
in two registers. Compute the sign of each permutation in the first register and then invert its phase only when the permutation is odd. Consequently, we obtain the quantum state
$
 \frac{1}{\sqrt{n!}}\sum_{\sigma\in S_n}(-1)^{{\rm sgn}(\sigma)}\ket{\sigma}\ket{\sigma(G)}.
$
Recall that ${\rm sgn}(\sigma) = 0$ if $\sigma$ is even, and ${\rm sgn}(\sigma) = 1$ otherwise.
By discarding the second register, we immediately obtain a certain quantum state, say,  $\chi$ in the first register. Note that, since $\pi$ is odd, if $\sigma$ is odd (even, resp.) then $\sigma\pi$ is even (odd, resp.).
Therefore, it follows that $\chi = \rho_\pi^-$ if $G$ is an ``YES" instance with the unique non-trivial automorphism $\pi$, and $\chi = \iota$ otherwise.
\end{proof}

We are now ready to present a polynomial-time reduction from GA to \QSCDFF . This concludes that \QSCDFF\ is computationally at least as hard as GA 
for infinitely-many input lengths $n$ (and thus in worst-case).

\begin{theorem}\label{reduction}
If there exist a polynomial $k$ and a polynomial-time quantum algorithm 
that solves $k$-\QSCDFF\ with non-negligible advantage, then 
there exists a polynomial-time quantum algorithm that solves GA in the worst case for infinitely-many input lengths $n$.
\end{theorem}
\begin{proof}
We first show that GA is polynomial-time Turing equivalent to UniqueGA$_{\it ff}$. Later, we give a polynomial-time Turing reduction 
from UniqueGA$_{\it ff}$ to \QSCDFF . 
By combining these two reductions, we can reduce GA to \QSCDFF . 
The reduction from GA to UniqueGA$_{\it ff}$ we define 
is similar to the one given by 
K{\" o}bler, Sch{\" o}ning, and Tor{\' a}n~\cite{KST93}, who presented a polynomial-time Turing reduction 
from GA to UniqueGA.
Their polynomial-time algorithm for GA makes queries to a given oracle that correctly represents UniqueGA on the promised inputs. 
This algorithm works correctly because all queries made by the algorithm satisfy the promise of UniqueGA; that is, every query is 
a graph of even number of nodes with either a unique non-trivial automorphism without any fixed point or 
no non-trivial automorphism at all. By a slight modification of their reduction, we can obtain a reduction from GA to UniqueGA$_{\it ff}$.
Furthermore, it is also possible to make our length parameter $n$ 
satisfy the specific equation $n=2(2n'+1)$, where $n'\in\nat$. As a result, we obtain the following lemma.

\begin{lemma}\label{GA2UGAFF}
UniqueGA$_{\it ff}$ is polynomial-time Turing equivalent to GA.
\end{lemma}

In fact, a stronger statement than Lemma \ref{GA2UGAFF} holds. 
When a Turing reduction to a promise  
problem makes only queries that satisfy the promise of the problem, 
this reduction is called {\it smart} \cite{GS88}. 
%% Deleted in 2nd submission
%Such a smart reduction is desirable for a security 
%reduction of a cryptosystem. 
The reduction from GA to UniqueGA given by K{\"o}bler, Sch{\"o}ning, and Tor{\'a}n~\cite{KST93} is indeed smart, and therefore  
 so is our reduction. For readability, 
we postpone the proof of Lemma \ref{GA2UGAFF} until Appendix.

{}From Lemma \ref{GA2UGAFF}, it suffices to construct a reduction from UniqueGA$_{\it ff}$ 
to \QSCDFF . Assume that there exist two polynomials $k$ and $p$ and also 
a polynomial-time quantum algorithm ${\cal A}$ such that, for infinitely many $n$'s, ${\cal A}$ 
solves $k$-\QSCDFF\ with advantage $1/p(n)$. 
Let us fix an arbitrary $n$ for which ${\cal A}$ solves $k$-\QSCDFF\ 
with advantage $1/p(n)$. 
On a given instance $G$ of UniqueGA$_{\it ff}$, we perform 
the following procedure:
\begin{description}
\magicwand
\item{(S1)} Generate from $G$ two sequences 
$S^+ = (\chi^{+ \otimes k},...,\chi^{+ \otimes k})$ and 
$S^- = (\chi^{- \otimes k},...,\chi^{- \otimes k})$ of $8p^2(n)n$ 
instances by running the generation algorithms given in 
Lemmas~\ref{positive-coset-sampling} and 
\ref{negative-coset-sampling}, respectively.
\item{(S2)} Invoke ${\cal A}$ on each component in $S^+$ and $S^-$ as an input.
Let 
$R^+ = ({\cal A}(\chi^{+ \otimes k}),...,{\cal A}(\chi^{+ \otimes k}))$ and 
$R^- = ({\cal A}(\chi^{- \otimes k}),...,{\cal A}(\chi^{- \otimes k}))$ be the resulting sequences of $8p^2(n)n$ entries.
\item{(S3)} Output YES if the difference $\ell$ 
between the number of 1's in $R^+$ and that in $R^-$ is at least $4p(n)n$;  
output NO otherwise.
\end{description}
Note that if $G$ is an ``YES" instance, then $S^{+}$ and $S^{-}$ should have the form  
$S^+ = (\rho_\pi^{+ \otimes k},...,\rho_\pi^{+ \otimes k})$ and $S^- = (\rho_\pi^{- \otimes k},...,\rho_\pi^{- \otimes k})$ of $8p^{2}(n)n$ entries; otherwise, we have  
$S^+ = S^- = (\iota^{\otimes k},...,\iota^{\otimes k})$.
%%%%%%
%%%%%%
\ignore{
$S^+ = (\overbrace{\rho_\pi^{+ \otimes k},...,\rho_\pi^{+ 
\otimes k}}^{8p^2(n)n})$ and $S^- = (\overbrace{\rho_\pi^{- 
\otimes k},...,\rho_\pi^{- \otimes k}}^{8p^2(n)n})$; otherwise, 
we have  
$S^+ = S^- = \overbrace{(\iota^{\otimes k},...,\iota^{\otimes k})}^{8p^2(n)n}$.  
}
%%%%%%%
%%%%%%%
Therefore, if $G$ is an ``YES" instance,  
the numbers of 1's in $R^+$ and in $R^-$ are highly likely different.  

\sloppy Finally, we estimate the difference $\ell$.  
Let $X^+$ and $X^-$ be two random variables respectively expressing the numbers of 1's in $R^+$ and in $R^-$. Assume that $G$ is an ``YES" instance. Since ${\cal A}$ solves $k$-\QSCDFF\ with advantage $1/p(n)$, we have $|\Pr[{\cal A}(\rho_\pi^{+ \otimes k})=1]-\Pr[{\cal A}(\rho_\pi^{- \otimes k})=1]| > 1/p(n)$. Next, we want to show that 
$
 \Pr[|X^+ - X^-| > 4p(n)n] > 1 - 2e^{-n}
$
using the H{\"o}ffding bounds, which are stated below. 

\begin{lemma}[H{\"o}ffding~\cite{Hoe63}]
Let $(X_1,...,X_m)$ be any sequence of
independent Bernoulli random variables on $\{0,1\}$ such that 
$\Pr[X_i=1] = p$ for any $i\in\{1,...,m\}$, and
let $X$ be a random variable expressing the number of 1's 
in the sequence, i.e., $X=\sum_{i=1}^m X_i$.
Then, for any $0\le\delta\le 1$, it holds that 
\[
 \Pr\left[X > (p+\delta)m \right] < e^{-2m\delta^2}\qquad\text{and}\qquad
 \Pr\left[X < (p-\delta)m \right] < e^{-2m\delta^2}. 
\]
\end{lemma}

For convenience, we define 
$p_L=\max\{\Pr[{\cal A}(\rho_\pi^{+ \otimes k})=1],\Pr[{\cal A}(\rho_\pi^{- \otimes k})=1]\}$ and 
$p_S=\min\{\Pr[{\cal A}(\rho_\pi^{+ \otimes k})=1],\Pr[{\cal A}(\rho_\pi^{- \otimes k})=1]\}$.
{}From our assumption, we obtain  $p_L-p_S>1/p(n)$. 
Note that $R^+$ and $R^-$ are precisely two sequences of 
$8p^2(n)n$ independent Bernoulli random variables on $\{0,1\}$ 
 with probabilities $p_L$ and $p_S$. We denote by 
$X_L$ ($X_S$, resp.) the number of 1's in the sequence associated 
with $p_L$ ($p_S$, resp.). The H{\"o}ffding bounds imply 
\[
 \Pr\left[X_L < (p_L-\delta)m \right] < e^{-n}\qquad\text{and}\qquad
 \Pr\left[X_S > (p_S+\delta)m \right] < e^{-n},
\]
where $m=8p^2(n)n$ and $\delta=1/(4p(n))$.
Since $p_L-p_S>1/p(n)$, we obtain $(p_L-p_S-2\delta)m > 4p(n)n$. 
{}From this inequality, it follows that 
\begin{eqnarray*}
 \Pr\left[|X^+ - X^-| > 4p(n)n \right]
 &\ge& \Pr\left[|X^+ - X^-| > (p_L-p_S-2\delta)m\right]\\
 &\ge& \Pr\left[X_L > (p_L-\delta)m\ \wedge\ X_S < (p_S+\delta)m\right].
\end{eqnarray*}
Since $X_L$ and $X_S$ are independent, we obtain a lower bound: 
\[
 \Pr\left[X_L > (p_L-\delta)m\ \wedge\ X_S < (p_S+\delta)m\right]
 \ge (1-e^{-n})^2 > 1 - 2e^{-n},
\]
from which we conclude that $\Pr[|X^+ - X^-| > 4p(n)n] > 1 - 2e^{-n}$.

Similarly, when $G$ is a ``NO" instance, we have 
$
 \Pr[|X^+ - X^-| < 4p(n)n] > 1 - 2e^{-n}
$.
This guarantees that the above procedure solves UniqueGA$_{\it ff}$ efficiently.
\end{proof}

%Next, we argue our second claim.  
As noted in Section~\ref{intro}, our distinction problem \QSCDFF\ has 
its roots in SHSP. A special case of SHSP is known to be reducible 
to \DIST, which is a problem of distinguishing between 
$\{\rho_\pi^+(n)\}_{n\in N}$ and 
$\{\iota(n)\}_{n\in N}$. As Hallgren, Moore, R{\"o}tteler, Russell, 
and Sen~\cite{HMRRS06} demonstrated, 
solving \DIST\  from $o(n \log{n})$ identical copies is 
impossible even for a time-unbounded quantum algorithm.
Now, we show a close relationship between \QSCDFF\ and \DIST.
%this distinction problem between $\{\rho_\pi^+(n)\}_{n\in N}$ and 
%$\{\iota(n)\}_{n\in N}$. 

Before stating our claim (Theorem \ref{SHSP}), we present an algorithm that converts 
$\rho_\pi^+$ to $\rho_\pi^-$ for each fixed $\pi\in {\cal K}_n$. This algorithm is a key to the proof of the theorem and further to the construction of a quantum PKC in the subsequent section.

\begin{lemma}[Conversion Algorithm]\label{rho-conversion}
There exists a polynomial-time quantum algorithm that, with certainty, converts  $\rho_\pi^+(n)$ into $\rho_\pi^-(n)$ and keeps $\iota(n)$ as it is for any parameter $n\in N$ and any hidden permutation $\pi\in{\cal K}_n$.
\end{lemma}
\begin{proof}
Let $n\in N$ be arbitrary. First, recall the definition of ${\rm sgn}(\sigma)$:
${\rm sgn}(\sigma)=0$ if $\sigma$ is even and ${\rm sgn}(\sigma)=1$ otherwise.
Let $\pi\in{\cal K}_n$ be any hidden permutation and consider its corresponding quantum state $\rho_\pi^+$. On input  $\rho_\pi^+$, our desired algorithm simply inverts its phase according to the sign of the permutation. This is done by performing the following transformation:
\[
 \ket{\sigma}+\ket{\sigma\pi} \longmapsto (-1)^{{\rm sgn}(\sigma)}\ket{\sigma} + (-1)^{{\rm sgn}(\sigma\pi)}\ket{\sigma\pi}.
\]
Note that determining the sign of a given permutation takes only  
time polynomial in $n$. 
Since $\pi$ is odd, ${\rm sgn}(\sigma)$ and ${\rm sgn}(\sigma\pi)$ are 
different; thus, 
the above algorithm obviously converts $\rho_\pi^+$ to $\rho_\pi^-$.
Moreover, the algorithm does not alter the quantum state $\iota$.
\end{proof}

%Hallgren et al.~\cite{HMRRS06} recently 
%proved it impossible to distinguish  
%between two certain coset states over the symmetric group
%even if $o(n\log{n})$ copies are provided. Their result implies that there is no quantum 
%algorithm distinguishing between $\{\rho_\pi^+(n)^{\otimes k(n)}\}_{n\in N}$ and $\{\iota(n)^{\otimes k(n)}\}_{n\in N}$ for $k(n)=o(n\log{n})$.

The intractability result of DIST \cite{HMRRS06}, stated above, 
also holds for \QSCDFF . {}To prove this claim, we want to 
show in Theorem \ref{SHSP} that \DIST\ can be reduced to
\QSCDFF\ in polynomial time. As a result,  
no time-unbounded quantum algorithm can solve \QSCDFF\ from $o(n\log{n})$ copies. The proof of
 the theorem requires a quantum version of a so-called {\em hybrid argument} used in computational cryptography.

\begin{theorem}\label{SHSP}
Let $k$ be any polynomial.
If there exists a quantum algorithm ${\cal A}$ such that 
\[
 \left|\Pr\limits_{\cal A}[{\cal A}(\rho_\pi^{+}(n)^{\otimes k(n)})=1]-\Pr\limits_{\cal A}[{\cal A}(\rho_\pi^{-}(n)^{\otimes k(n)})=1]\right| > \varepsilon(n)
\]
for any security parameter $n\in N$, then there exists a quantum algorithm ${\cal B}$ 
such that, for each $n\in N$,  
\[
 \left|\Pr\limits_{\cal B}[{\cal B}(\rho_\pi^{+}(n)^{\otimes k(n)})=1]-\Pr\limits_{\cal B}[{\cal B}(\iota(n)^{\otimes k(n)})=1]\right| > \frac{\varepsilon(n)}{4}.
\]
\end{theorem}
\begin{proof}
Fix $n\in N$ arbitrarily and we hereafter omit this parameter $n$. Assume that a quantum algorithm ${\cal A}$ distinguishes between 
$\rho_\pi^{+ \otimes k}$ and $\rho_\pi^{- \otimes k}$ with advantage at least 
$\varepsilon(n)$.
Let ${\cal A}'$ be the algorithm that applies the conversion algorithm of 
Lemma~\ref{rho-conversion}
to a given state $\chi$ (which is either $\rho_\pi^{+ \otimes k}$ or $\iota^{\otimes k}$) 
and then feeds the 
resulting state $\chi'$ (either $\rho_\pi^{- \otimes k}$ or $\iota^{\otimes k}$) to ${\cal A}$. 
It thus follows that ${\cal A}'(\rho_\pi^{+ \otimes k})={\cal A}(\rho_\pi^{- \otimes k})$ and ${\cal A}'(\iota^{\otimes k})={\cal A}(\iota^{\otimes k})$. By the triangle inequality, we have 
\[
  \left|\Pr\limits_{\cal A}[{\cal A}(\rho_\pi^{+ \otimes k})=1] - 
\Pr\limits_{\cal A}[{\cal A}(\iota^{\otimes k})=1] \right|
+ \left|\Pr\limits_{{\cal A}'}[{\cal A}'(\rho_\pi^{+ \otimes k})=1] - 
\Pr\limits_{{\cal A}'}[{\cal A}'(\iota^{\otimes k})=1] \right|
> \varepsilon(n)
\]
for any parameter $n\in N$. This inequality leads us to either 
\[
 \left|\Pr\limits_{\cal A}[{\cal A}(\rho_\pi^{+ \otimes k})=1] 
- \Pr\limits_{\cal A}[{\cal A}(\iota^{\otimes k})=1] \right| > \frac{\varepsilon(n)}{2}
\]
or
\[
 \left|\Pr\limits_{{\cal A}'}[{\cal A}'(\rho_\pi^{+ \otimes k})=1] 
- \Pr\limits_{{\cal A}'}[{\cal A}'(\iota^{\otimes k})=1] \right| > \frac{\varepsilon(n)}{2}.
\]
To complete the proof, we design the desired algorithm ${\cal B}$ as follows: first choose either  ${\cal A}$ or ${\cal 
A}'$ at random and then simulate the chosen algorithm.
It is easy to verify that ${\cal B}$ distinguishes between $\rho_\pi^{+ \otimes k}$ and $\iota^{\otimes k}$ with advantage at least $\varepsilon(n)/4$. 
\end{proof}

%%%%%%%%%%%%%%%%%%%%%%%%%%%%%%%%%%%%%%%%%%%%%%%%%%%%%%%%%%%%%%%%%%%%
\section{An Application to a Quantum Public-Key Cryptosystem}\label{sec:application}
%%%%%%%%%%%%%%%%%%%%%%%%%%%%%%%%%%%
Section~\ref{sec:property} has shown the three useful cryptographic properties of \QSCDFF . Founded on these properties, we wish to construct a quantum PKC whose security is guaranteed by the computational hardness of \QSCDFF\ (which can be further reduced to the hardness of GA). 
As the first step, we give an efficient quantum algorithm that generates $\rho_\pi^+$ from $\pi$.

\begin{lemma}[$\boldsymbol{\rho_\pi^+}$-Generation Algorithm]\label{rho-generation}
There exists a polynomial-time quantum algorithm that, on input  $\pi\in {\cal K}_n$, generates the quantum state $\rho_\pi^+$ with probability $1$.
\end{lemma}
\begin{proof}
The desired generation algorithm, which is given below, uses two 
registers. Here, we omit the proof of the correctness of the given algorithm because the correctness is obvious from the description of the algorithm.
\begin{description}
\magicwand
\item[{\rm (G1)}] Prepare the state $\ket{0}\ket{id}$ in two quantum registers.
\item[{\rm (G2)}] Apply the Hadamard transformation to the first register to obtain the state 
$
 \frac{1}{\sqrt{2}}(\ket{0}+\ket{1})\ket{id}.
$
\item[{\rm (G3)}] Perform the Controlled-$\pi$ on the both registers and we then obtain the state  
$
 \frac{1}{\sqrt{2}}(\ket{0}\ket{id}+\ket{1}\ket{\pi}).
$
\item[{\rm (G4)}] Subtract $1$ from the content of the first register  only  when the second register contains $\pi$. This  process gives rise to the state 
$
 \frac{1}{\sqrt{2}}(\ket{0}\ket{id}+\ket{0}\ket{\pi}).
$
\item[{\rm (G5)}] Apply a uniformly random permutation $\sigma$ to 
the content of the second register from the left. The whole quantum 
system then becomes  
$
 \frac{1}{\sqrt{2}}(\ket{0}\ket{\sigma}+\ket{0}\ket{\sigma\pi}).
$
\item[{\rm (G6)}] Output the content of the second register, which produces the state $\rho_{\pi}^{+}$ with probability $1$.
\end{description}
\end{proof}

Hereafter, we describe our quantum PKC and then give its security proof. 
For the security proof, in particular, we need to clarify our model of adversary's attack. Of all attack models discussed in \cite{BDPR98}, we use  
a quantum analogue of {\em the indistinguishability against the chosen 
plaintext attack (IND-CPA)}. Our scenario is precisely as follows:

\begin{quote}
Suppose that large-scale quantum and classical networks 
connect a unique network administrator, acting as a trusted 
third party, and numerous ``ordinary'' network users, 
some of who might possibly be malicious against other users. 
These parties are all capable of running polynomial-time quantum 
algorithms.
In particular, the administrator (say, Charlie) can 
communicate with each 
network user via a secure, authenticated communication channel; 
namely, he can deliver to each individual user a piece of 
information (both quantum and classical bits) 
correctly and securely through this channel. 
It is most likely that a financial reason could force ordinary users to 
rely on cheap but insecure channels for daily person-to-person communication 
with other users. {}From such an insecure channel, a 
malicious party (say, Eve) might wiretap the communication. 
To ensure user's    
secure communication, upon a request from a user (say, Bob) 
who wants to receive a message from other users, 
Charlie generates a decryption (or private) key $\pi$ 
and sends it through the secure channel to Bob.  
Charlie also generates an encryption (or public) key $\rho_{\pi}^{+}$ 
for anyone who wants to communicate with Bob. 

Now, suppose that a honest party, called Alice, 
wishes to send Bob a classical single-bit message securely. 
For this purpose, she first requests Charlie 
for Bob's encryption key $\rho_{\pi}^{+}$. Using this key, 
she encrypts her secret message into a quantum state 
$\rho$ (either $\rho_{\pi}^{+}$ or $\rho_{\pi}^{-}$)  as a ciphertext   
and then sends it to Bob through an available insecure quantum channel.  
To eavesdrop Alice's secret message, 
Eve intercepts Alice's ciphertext $\rho$. 
In addition, since Eve is also a legitimate network user, 
she can request numerous copies of 
the encryption key $\rho_{\pi}^{+}$ from Charlie
(within a polynomial amount of time). 
Finally, Eve attempts to learn the information involved with 
Alice's secret message 
by applying a certain polynomial-time quantum algorithm to  
the ciphertext $\rho$ as well as  a polynomially many 
copies of the encryption key $\rho_{\pi}^{+}$ obtained from Charlie as supplemental information. 
\end{quote}

In the case of classical 
chosen plaintext attack, all that Eve can collect 
are Alice's ciphertext and Bob's encryption key.  
Our scenario is a natural generalization of this classical 
case because  
Eve obtains only a quantum state representing 
Alice's encrypted message and copies of a quantum 
state serving as an encryption key. 

Our scenario demands that the administrator should generate and 
distribute user's private and public keys. In a practical 
framework of classical PKCs, such a scenario has been frequently used; 
for example, a governmental agency may be authorized as a third party to handle those user's keys. 
Note that Charlie's distribution of decryption keys is done 
through the secure channel only once at the key setup. 
With their own single 
decryption keys, all the users can transmit their messages securely to others
a reasonably large number of times, even without 
any extra secret information shared among them. To the contrary, SKCs 
require the users to share symmetric secret keys between every pair of them.
Thus, even under this scenario, we can enjoy advantages of PKCs over 
SKCs that stem from the asymmetry of keys in many-to-many communication.

Now, we explain our quantum PKC protocol in detail. In 
our protocol, Alice transmits a single-bit message to Bob using an  $O(n\log n)$-qubit-long encryption key. 
Our protocol consists of three phases: 
 key setup phase, key transmission phase, and message transmission phase. Figure 1 illustrates our protocol. 
%
%\vspace*{16mm}
%
%\begin{center}
%------------\\
%Figure~1\\
%------------
%\end{center}
%
%\vspace*{16mm}

\begin{center}
\mbox{}\\
\resizebox{7cm}{!}{ 
\includegraphics*{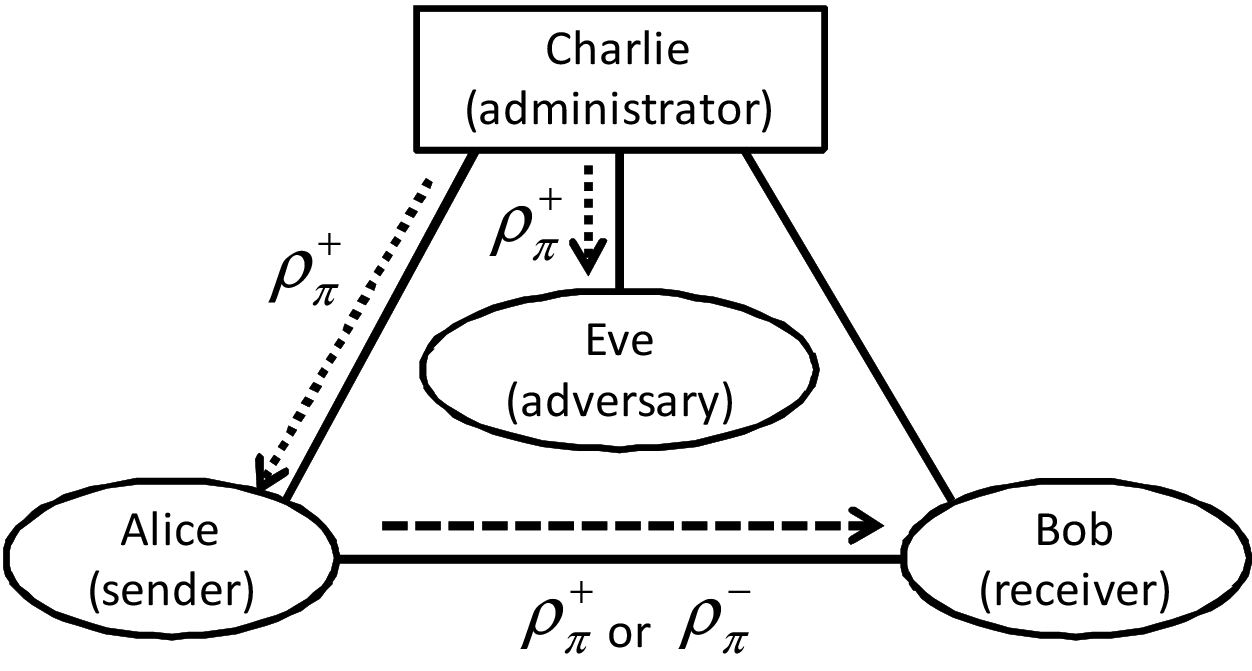}
}\\
Figure~1: our public-key cryptosystem
\end{center}
The following is the step-by-step description of 
our quantum PKC protocol.\\
\noindent
{\sf [Key setup phase]}\vspace{-2mm}
\begin{description}
\magicwand
\item[{\rm (A1)}] Charlie generates Bob's decryption 
key $\pi$ uniformly at random from ${\cal K}_n$, and then sends it to Bob via a secure and authenticated channel. 
\end{description}\vspace{-2mm}
\noindent
{\sf [Key transmission phase]}\vspace{-2mm}
\begin{description}
\magicwand
\item[{\rm (A2)}] Alice requests Bob's encryption key from Charlie.
\item[{\rm (A3)}] Using $\pi$, Charlie generates a copy of the encryption key $\rho_\pi^+$.
\item[{\rm (A4)}] Alice obtains a copy of the encryption key $\rho_{\pi}^{+}$ from Charlie.
\end{description}\vspace{-2mm}
{\sf [Message transmission phase]} \vspace{-2mm}
\begin{description}
\magicwand
\item[{\rm (A5)}] Alice encrypts $0$ or $1$ respectively 
into $\rho_\pi^+$ or $\rho_\pi^-$  and then sends this 
encrypted message to Bob.
\item[{\rm (A6)}] Bob decrypts Alice's message using the decryption key $\pi$.
\end{description}
Step (A1) can be implemented as follows. Recall that $\pi\in {\cal K}_n$ consists of $n/2$ disjoint transpositions. We first choose 
distinct two numbers $i_1$ and $i_2$ from $\{1,2,...,n\}$ uniformly at random, 
and make a transposition $(i_1\ i_2)$. Next, choosing other distinct two 
numbers $i_3$ and $i_4$ from $\{1,2,...,n\}\setminus\{i_1,i_2\}$ uniformly 
at random, we make another transposition $(i_3\ i_4)$. 
By repeating this process, $n/2$ disjoint transpositions are chosen 
uniformly at random. {}From them, define $\pi = (i_1,i_2)\cdots (i_{n/2-1},i_{n/2})$. 
Step (A3) is done by the $\rho_\pi^+$-generation algorithm
of Lemma~\ref{rho-generation}. The conversion algorithm of Lemma~\ref{rho-conversion} implements Step (A5) since
 Alice sends Bob either the received state $\rho_\pi^+$ or 
its converted state $\rho_\pi^-$.  
Finally, the distinguishing algorithm of
Theorem~\ref{trapdoor} implements Step (A6).

The security proof of our PKC is done by reducing GA to 
Eve's attacking strategy during 
the message transmission phase. Our reduction is a simple modification of the reduction given in Theorem~\ref{reduction}.

\begin{proposition}
Let ${\cal A}$ be any polynomial-time quantum adversary who attacks our quantum PKC during the message transmission phase. Assume that there exist two polynomials $p(n)$ and $l(n)$ satisfying that 
\[
 \left|
 \Pr\limits_{\pi, {\cal A}}[{\cal A}(\rho_\pi^+,\rho_\pi^{+ \otimes l(n)})=1]
-\Pr\limits_{\pi, {\cal A}}[{\cal A}(\rho_\pi^-,\rho_\pi^{+ \otimes l(n)})=1]
 \right|
 > \frac{1}{p(n)}
\]
for infinitely many parameters $n\in N$. Then, there exists a polynomial-time quantum algorithm that solves GA  for infinitely many input sizes $n$ in the worst case with non-negligible probability.
\end{proposition}
\begin{proof}
The proposition immediately follows from the proof of Theorem~\ref{reduction} by replacing $\rho_\pi^{+ \otimes k}$, $\rho_\pi^{- \otimes k}$,  and $\iota^{\otimes k}$ in the proof with $(\rho_\pi^+,\rho_\pi^{+ \otimes l(n)})$, $(\rho_\pi^-,\rho_\pi^{+ \otimes l(n)})$, and $(\iota,\iota^{\otimes l(n)})$, respectively.
\end{proof}

%%%%%%%%%%%%%%%%%%%%%%%%%%%
%%%%%%%%%%%%%%%%%%%%%%%%%%%
\section{A Generalization of \QSCDFF}

In our \QSCDFF -based quantum PKC, Alice encrypts a single-bit message using an $O(n\log n)$-qubit encryption key. We wish to show how to increase the size of Alice's encryption message and construct a multi-bit 
quantum PKC built upon a generalization of \QSCDFF , called \QSCDC\ (QSCD with cyclic permutations), which is a distinction problem
among {\em multiple ensembles} of quantum states. 
Recall that Definition~\ref{q-ind} has introduced the notion of computational indistinguishability between two  
ensembles of quantum states. This notion can be naturally generalized as follows to multiple quantum state ensembles.
\begin{definition}
We say that $m$ ensembles 
$\{\rho_0(l)\}_{l\in \nat},...,\{\rho_{m-1}(l)\}_{l\in \nat}$ of quantum states are 
{\em computationally indistinguishable\/} 
if, for any distinct pair $i,j\in\integer_{m}$, the advantage of  
distinguishing between the two ensembles
$\{\rho_{i}(l)\}_{l\in \nat}$ and $\{\rho_{j}(l)\}_{l\in \nat}$
is negligible for any polynomial-time 
quantum algorithm ${\cal A}$; namely, for any two ensembles 
$\{\rho_{i}(l)\}_{l\in \nat}$ and $\{\rho_{j}(l)\}_{l\in \nat}$,
any polynomial $p$, any 
polynomial-time quantum algorithm ${\cal A}$, and any sufficiently large number $l$, it holds that 
\[
 \left| \Pr_{\cal A}[{\cal A}(\rho_{i}(l))=1] - \Pr_{\cal A}[{\cal A}(\rho_{j}(l))=1] \right| < \frac{1}{p(l)}.
\]
The distinction problem among the ensembles 
$\{\rho_0(l)\}_{l\in \nat},...,\{\rho_{m-1}(l)\}_{l\in \nat}$ is 
said to be {\em solvable with non-negligible advantage\/} if 
the ensembles are not computationally indistinguishable; that is,  
there exist two ensembles $\{\rho_{i}(l)\}_{l\in\nat}$ and 
$\{\rho_{j}(l)\}_{l\in\nat}$, a polynomial-time quantum algorithm 
${\cal A}$, and a polynomial $p$ such that 
\[
 \left| \Pr_{\cal A}[{\cal A}(\rho_{i}(l))=1] - \Pr_{\cal A}[{\cal A}(\rho_{j}(l))=1] \right| > \frac{1}{p(l)}
\]
for infinitely many numbers $l\in\nat$.
\end{definition}

We wish to define a specific distinction problem, denoted succinctly  \QSCDC ,  among $m$ ensembles of quantum states.
First, we define a new hidden permutation, which will be encoded into 
 certain quantum states. 
For any fixed number $n\in\nat$, let us assume that $m\geq 2$ and $m$ divides $n$.
The new hidden permutation $\pi$ consists of 
disjoint $n/m$ cyclic permutations of length $m$; in other words, 
$\pi$ is  of the form 
\[
 \pi = (i_0\, i_1\, \cdots\, i_{m-1})\cdots
       (i_{n-m}\, i_{n-m+1}\, \cdots\, i_{n-1}),
\]
where $i_0,...,i_{n-1}\in\integer_n$ and $i_s\neq i_t$ 
if $s\neq t$ for any pair $(s,t)$.
Such a permutation $\pi$ has the following two properties:
($i$) $\pi$ has no fixed points (i.e., $\pi(i)\neq i$ for any $i\in\integer_n$) and 
($ii$) $\pi$ is of order $m$ (i.e., $\pi^m= id$). For convenience, 
we denote by ${\cal K}^m_n$ ($\subseteq S_n$) the set of all such permutations.

With a help of the hidden permutation $\pi$, we can 
define the new quantum states $\ket{\Phi_{\pi,s}^\sigma}$ as follows.
For each $\sigma\in S_n$, $\pi\in {\cal K}_n^m$, and $s\in \integer_m$, let 
\[
 \ket{\Phi_{\pi,s}^\sigma} = \frac{1}{\sqrt{m}}\sum_{t=0}^{m-1}\omega_m^{st}\ket{\sigma\pi^t},
\]
where $\omega_m= e^{2\pi i/m}$. 
At last, the distinction problem \QSCDC\ is defined in the following way.

\begin{definition}
The problem \QSCDC\ is a distinction problem among $m$ 
ensembles $\{\rho_\pi^{(0)}(n)^{\otimes k(n)}\}_{n\in \nat},$
$..., \{\rho_\pi^{(m-1)}(n)^{\otimes k(n)}\}_{n\in \nat}$ of quantum states, where $k$ is an arbitrary polynomial and the notation
$\rho_\pi^{(s)}(n)$ denotes the mixed state 
$\frac{1}{n!}\sum_{\sigma\in S_n}\ketbra{\Phi_{\pi,s}^\sigma}{\Phi_{\pi,s}^\sigma}$
for each $\pi\in {\cal K}^m_n$. 
When $k$ is fixed, we use the notation $k$-\QSCDC\ instead.
\end{definition}

Similar to the case of \QSCDFF , we also drop the parameter $n$ wherever possible. 
Note that \QSCDFF\ coincides with \QSCDC\ with
$m=2$ and $n$ is of the form $2(2n'+1)$ for a certain number $n'\in\nat$.

This new problem \QSCDC\ also enjoys useful cryptographic properties.
We first present a trapdoor of \QSCDC . 
In the case of \QSCDFF , because its trapdoor information $\pi$ 
is a permutation of order two, we encode only a single bit into the  both quantum states $\rho_\pi^{+}$ and $\rho_\pi^{-}$.  On the contrary, since  
\QSCDC\ uses a permutation $\pi$ of order $m\geq 2$, it is possible to encode 
$\log{m}$ bits into the $m$ quantum states $\rho_\pi^{(0)},...,\rho_\pi^{(m-1)}$. 

Now, we present a generalized distinguishing algorithm working for  $\rho_\pi^{(s)}$'s.

\begin{theorem}[Generalized Distinguishing Algorithm]\label{generalized-decryption}
There exists a polynomial-time quantum algorithm that, for each $n\in\nat$, $\pi\in {\cal K}^m_n$, and $s\in\integer_{m}$, decrypts $\rho_\pi^{(s)}(n)$ to $s$ with  exponentially-small error 
probability. 
\end{theorem}
\begin{proof}
Let $\chi$ be any given quantum state of the form $\rho_\pi^{(s)}$ 
for a certain 
hidden permutation $\pi\in{\cal K}_n^m$ and also a certain 
hidden parameter $s$.
Note that $\chi$ is a mixture of all pure states $\ket{\Phi_{\pi,s}^\sigma}$ 
over a randomly chosen $\sigma\in S_n$. It thus suffices to 
give a polynomial-time quantum algorithm that decrypts   
$\ket{\Phi_{\pi,s}^\sigma}$ to $s$ for each fixed $\sigma$.
Such an algorithm can be given by conducting the following {\em Generalized Controlled-$\pi$ 
Test}, which is a straightforward generalization of the distinguishing algorithm given in the proof of Theorem~\ref{trapdoor}. 
To define this test, we first recall 
the quantum Fourier transformation $F_m$ over $\integer_m$ as well as 
its inverse $F_m^{-1}$: for any $x\in\integer_m$, 
\[
 F_m\ket{x} = \frac{1}{\sqrt{m}}\sum_{y\in\integer_m} \omega_m^{xy} \ket{y}
\quad\text{and}\quad
 F_m^{-1}\ket{x} = \frac{1}{\sqrt{m}}\sum_{y\in\integer_m} \omega_m^{-xy} \ket{y}.
\]
The Generalized Controlled-$\pi$ Test is described below.\\

\noindent{\sf [Generalized Controlled-$\pi$ Test]}\vspace{-2mm}
\begin{description}
\magicwand
\item[{\rm (D1')}] Prepare two quantum registers. The first register holds a 
control string, initially set to $\ket{0}$, and the second register  holds the quantum state 
$\ket{\Phi_{\pi,s}^\sigma}$. Apply the inverse Fourier transformation $F_m^{-1}$ to the 
first register. 
Meanwhile, assume that we can perform the 
Fourier transformation exactly. The entire system then becomes 
\[
 \frac{1}{\sqrt{m}}\sum_{r=0}^{m-1}\ket{r}\ket{\Phi_{\pi,s}^\sigma}
=\frac{1}{m}\sum_{r,t}\omega_m^{st}\ket{r}\ket{\sigma\pi^t}.
\]

\item[{\rm (D2')}] Apply $\pi$ to the content of the second register  $r$ times from the right. The state of the entire system evolves into  
\[
 \frac{1}{m}\sum_{r,t}\omega_m^{st}\ket{r}\ket{\sigma\pi^{r+t \bmod m}}.
\]
\item[{\rm (D3')}] Apply the Fourier transformation $F_m$ to the first register and we then obtain the state
\begin{eqnarray*}
& & \frac{1}{m}\sum_{r,t}\frac{1}{\sqrt{m}}\sum_{r'=0}^{m-1}\omega_m^{rr'}\ket{r'}\omega_m^{st}\ket{\sigma\pi^{r+t \bmod m}}\\
&=& \frac{1}{m^{3/2}}\sum_{r,r',t}\omega_m^{st+rr'}\ket{r'}\ket{\sigma\pi^{r+t \bmod m}}\\
&=& \frac{1}{m^{3/2}}\sum_{r,t}\omega_m^{s(r+t)}\ket{s}\ket{\sigma\pi^{r+t \bmod m}}
+   \frac{1}{m^{3/2}}\sum_{r,t,r'\neq s}\omega_m^{st+rr'}\ket{r'}\ket{\sigma\pi^{r+t \bmod m}}\\
&=& \frac{1}{\sqrt{m}}\sum_{u}\omega_m^{su}\ket{s}\ket{\sigma\pi^{u}}
+   \frac{1}{m^{3/2}}\sum_{r,u,r'\neq s}\omega_m^{su+r(r'-s)}\ket{r'}\ket{\sigma\pi^{u}}\quad\text{$(u:=r+t \bmod m)$}\\
&=& \frac{1}{\sqrt{m}}\sum_{u=0}^{m-1}\omega_m^{su}\ket{s}\ket{\sigma\pi^{u}}
\;\;=\;\; \ket{s}\ket{\Phi_{\pi,s}^\sigma} \quad(\text{since $\sum_r \omega_m^{su+r(r'-s)}=0$ for any $u,s,r'(\ne s)$}).
\end{eqnarray*}
\item[(D4')] Finally, measure the first register in the computational basis and output 
the measured result $s$ in $\integer_m$.
\end{description}
The error probability of the above algorithm depends only on the precision of the 
Fourier transformation over $\integer_m$. As shown in \cite{Kit95},  the quantum Fourier transformation can be implemented  
with exponentially-small error probability
by an application of the approximated quantum Fourier transformation. 
Therefore, the theorem follows.
\end{proof}

Similar to \QSCDFF ,
the average-case hardness of \QSCDC\ coincides with its 
worst-case hardness.
\begin{theorem}
Let $k$ be any polynomial.
Assume that there exists a polynomial-time quantum algorithm ${\cal A}$
that solves $k$-\QSCDC\ with non-negligible advantage 
for a uniformly random permutation 
$\pi\in{\cal K}_n^m$; namely, there exist two numbers $s,s'\in\integer_{m}$ and 
a polynomial $p$ such that, for infinitely many numbers $n\in \nat$,
\[
 \left| \Pr\limits_{\pi, {\cal A}}[{\cal A}(\rho_\pi^{(s)}(n)^{\otimes k(n)}) = 1] 
      - \Pr\limits_{\pi, {\cal A}}[{\cal A}(\rho_\pi^{(s')}(n)^{\otimes k(n)}) = 1] \right| > \frac{1}{p(n)},
\]
where $\pi$ is chosen uniformly at random from ${\cal K}_n^m$.
Then, there exists a polynomial-time quantum algorithm ${\cal B}$ that solves 
$k$-\QSCDC\ with non-negligible advantage.
\end{theorem}
\begin{proof}
This proof follows 
an argument in the proof of Theorem~\ref{wa-reduction}. 
Here, we give only a sketch of our desired algorithm ${\cal B}$. 
Choose a uniformly random permutation $\tau\in S_n$ and then apply it to 
$\ket{\Phi_{\pi,s}^\sigma}$ from the right. Now, we obtain the state
\[
  \frac{1}{\sqrt{m}}\sum_{t=0}^{m-1} \omega_m^{st}\ket{\sigma\pi^{t} \tau} 
= \frac{1}{\sqrt{m}}\sum_{t=0}^{m-1} \omega_m^{st}\ket{\sigma\tau \tau^{-1}\pi^{t} \tau}
= \frac{1}{\sqrt{m}}\sum_{t=0}^{m-1} \omega_m^{st}\ket{\sigma\tau (\tau^{-1}\pi\tau)^t}.
\]
Note that $\rho_{\tau^{-1}\pi\tau}^{(s)}(n) = \frac{1}{n!}\sum_{\sigma\in S_n} \ketbra{\Phi_{\tau^{-1}\pi\tau,s}^{\sigma\tau}}{\Phi_{\tau^{-1}\pi\tau,s}^{\sigma\tau}}$ 
is an average-case instance of \QSCDC\ 
since $\tau^{-1}\pi \tau$ is distributed uniformly at random over 
${\cal K}_n^m$. Finally, apply the average-case algorithm ${\cal A}$.  
\end{proof}

We will exhibit a quantum algorithm that generates the quantum state $\rho_\pi^{(s)}$ efficiently from $\pi$ and $s$.
This generation algorithm will be used to generate encryption keys in our \QSCDC -based multi-bit quantum PKC.

\begin{lemma}[$\boldsymbol{\rho_\pi^{(s)}}$-Generation Algorithm]\label{rho-generation2}
There exists a polynomial-time quantum algorithm that generates 
$\rho_\pi^{(s)}$ for any $s\in\integer_{m}$
and any $\pi\in{\cal K}_n^m$ with exponentially-small error probability.
\end{lemma}
\begin{proof}
The desired algorithm is a straightforward generalization of the 
$\rho_\pi^+$-generation algorithm given in the proof of Lemma \ref{rho-generation}. 
Using the approximated Fourier 
transformation \cite{Kit95} instead of the Hadamard transformation,  we can efficiently approximate from  $\pi$ 
the Fourier transformation $F_\pi$ over the cyclic group $\{id,\pi,\pi^2,...,\pi^{m-1}\}$:  
\[
 F_\pi\ket{\pi^s} = \frac{1}{\sqrt{m}}\sum_{t=0}^{m-1}\omega_m^{st}\ket{\pi^t}
\]
by employing an argument similar to the proof of Lemma~\ref{rho-generation}.
Hence, we can 
perform $ F_\pi$ on $\ket{\pi^s}$ with exponentially-small error probability. 

Since the initial state $\ket{\pi^s}$ can be easily generated from  $\pi$,
we immediately obtain an efficient approximation of $F_\pi\ket{\pi^s}$.
By applying a uniformly-random permutation $\sigma\in S_n$ to the resulting state
from the left, the desired state $\rho_\pi^{(s)}$ can be obtained 
with exponentially-small error probability.
\end{proof}

Toward the end of this section, we present our multi-bit quantum PKC, based on \QSCDC . \\
\noindent
{\sf [Key setup phase]}\vspace{-2mm}
\begin{description}
\magicwand
\item[{\rm (A1')}] As Bob's decryption key, Charlie chooses an element   $\pi$ uniformly at random from ${\cal K}_n$ and then sends it to Bob 
via a secure, authenticated channel. 
\end{description}\vspace{-2mm}
\noindent
{\sf [Key transmission phase]}\vspace{-2mm}
\begin{description}
\magicwand
\item[{\rm (A2')}] Alice requests Bob's encryption key from Charlie.
\item[{\rm (A3')}] Charlie generates a copy of the encryption key
$(\rho_\pi^{(0)},...,\rho_\pi^{(m-1)})$ from $\pi$ and sends it to Alice.
\item[{\rm (A4')}] Alice receives this copy of the encryption key 
from Charlie.
\end{description}\vspace{-2mm}
{\sf [Message transmission phase]} \vspace{-2mm}
\begin{description}
\magicwand
\item[{\rm (A5')}] If her message is $s\in\integer_{m}$,  
Alice picks up $\rho_\pi^{(s)}$. She sends it to Bob as a ciphertext.

\item[{\rm (A6')}] Bob decrypts Alice's message using the decryption key $\pi$.
\end{description}
By choosing cycles one by one sequentially, we can perform Step (A1'). 
The $\rho_\pi^{(s)}$-generation algorithm
of Lemma~\ref{rho-generation2} immediately implements Step (A3'). 
Alice can encrypt her message $s$ simply by choosing 
$\rho_\pi^{(s)}$ out of 
the series $(\rho_\pi^{(0)},...,\rho_\pi^{(m-1)})$.
Finally, the generalized distinguishing algorithm in 
Theorem~\ref{generalized-decryption} achieves Step (A6').

As the final remark, we refer to a drawback of the above multi-bit encryption scheme. A major drawback is that Charlie should send 
Alice all the series $(\rho_\pi^{(0)},...,\rho_\pi^{(m-1)})$ as 
Bob's encryption key, simply because of the lack of a sophisticated converting algorithm among different encryption keys without knowing the hidden decryption key $\pi$. 
This \QSCDC -based encryption scheme requires an $O(mn\log n)$-qubit encryption key to encrypt a $\log{m}$-bit message whereas the \QSCDFF -based encryption scheme needs an $O(n\log n)$-qubit key per a $1$-bit message. In a quick comparison, there seems to be no advantage of 
the \QSCDC -based scheme over the \QSCDFF -based scheme 
in terms of the ratio between 
message length and encryption key length. 

This drawback stems from the conversion algorithm, given in Lemma \ref{rho-conversion}, used to swap $\rho_{\pi}^{+}$ and $\rho_{\pi}^{-}$ 
in the \QSCDFF -based single-bit encryption scheme. 
This conversion algorithm utilizes   
the ``parity'' of permutations $\sigma$ and $\sigma\pi$ to invert 
their phases without using any information on $\pi$. More precisely,
the algorithm implements the homomorphism $f$ from $S_n$ to 
$\{+1,-1\}$ ($\cong \integer/2\integer$) 
satisfying that $f(\sigma)=+1$ ($-1$, resp.) if $\sigma$ is even (odd, resp.). 
Unfortunately, the same algorithm {\em fails} for \QSCDC\ 
because no homomorphism maps $S_n$ to 
$\{1,\omega_m,...,\omega_m^{m-1} \}$ ($\cong \integer/m\integer$). This is shown as follows. 
Let us assume, to the contrary, that there exists a homomorphism 
$g$ mapping $S_n$ to $\{1,\omega_m,...,\omega_m^{m-1} \}$.
The {\em fundamental homomorphism theorem} implies that  
$S_n/\Ker(g) \cong \integer/m\integer$;
namely, there exists an isomorphism from $\sigma\Ker(g)$ to $g(\sigma)$
for every $\sigma\in S_n$. Note that $\Ker(g)$ is a normal subgroup in $S_n$. It is known that such a normal subgroup in $S_n$ equals either 
the trivial group $\{id\}$ or the alternation group 
$A_n = \{\sigma\in S_n:\ {\rm sgn}(\sigma)=0\}$ 
since $A_n$ is a simple group for $n\ge 5$ (see, e.g., 
Theorem 3.2.1 in \cite{Rob95}).
Apparently, there is neither isomorphism between 
$\{\sigma A_n: \sigma\in S_n\}$ and $\integer/m\integer$ 
nor isomorphism between $\{\sigma: \sigma\in S_n\}$ 
and $\integer/m\integer$ if $n>4$ and $n\geq m > 2$. 
This contradicts our assumption on $g$. 

%%%%%%%%%%%%%%%
\section{Concluding Remarks}

We have shown that the computational distinction problem 
\QSCDFF\ satisfies quite useful cryptographic  
properties, which help us to design a quantum PKC whose security is 
guaranteed by the computational intractability of GA. 
Although GA is reducible to \QSCDFF\ in polynomial time, 
there seems to be a large gap between the hardness of GA and that of \QSCDFF\ because, in the proof of Theorem \ref{reduction}, all  combinatorial structures of an input graph for GA are completely lost in  constructing associated quantum states for \QSCDFF\ and, from such states, it is impossible to recover the original graph. It is therefore pressing to find a much better classical problem (for instance, the problems of finding a centralizer or finding a normalizer \cite{Luk93}) that almost matches the computational hardness of \QSCDFF . Since no fast quantum algorithm is known for \QSCDFF , discovering such a fast  algorithm for \QSCDFF\ may require new tools and novel proof techniques in quantum complexity theory. Besides our quantum states $\{\rho^+_{\pi}(n),\rho^{-}_{\pi}(n)\}$ used in \QSCDFF , it is imperative to 
continue searching for other pairs of ``simple'' quantum states whose computational indistinguishability is helpful to construct a more secure cryptosystem. 

Similar to \QSCDFF , \QSCDC\ also owns useful cryptographic 
properties, for which we have built a multi-bit quantum PKC. Throughout our study, it is not yet clear how difficult \QSCDC\ is and how secure  
our multi-bit quantum PKC truly is. If one successfully proves that the worst-case hardness of \QSCDC\ is lower-bounded by, e.g., 
the hardness of GA, then our multi-bit quantum PKC might find a more practical use in return.

\paragraph{Acknowledgments.} 
The authors are grateful to Hirotada Kobayashi and Claude Cr\'{e}peau for fruitful discussions, to John Watrous for useful comments on key ideas,
to Donald Beaver, Louis Salvail, and the anonymous reviewers of EUROCRYPT 2005 
and Journal of Cryptology for their valuable suggestions. 
The authors' thanks also go to 
Cristopher Moore for providing references to a historical account of
hidden subgroup problems. 
This research was partially supported by Grant-in-Aid for Young Scientists (B) No.17700007 (2005), Grant-in-Aid for Scientific Research on Priority Areas No.16092206 (2005), No.18300002 (2006), and No.21300002 (2009) from the Ministry of Education, Science, Sports and Culture. 

%%%%%%%%%%%%%%%%%%%%%%%%%%%%%%%%%%%%%%%%%%%%%%%%%%%%
%\bibliographystyle{abbrv}
%\bibliography{mybib}

%%%%%%%%%%%%%%%%%%%%%%%%%%%%%%%%
\section*{Appendix: A Reduction from GA to UniqueGA$_{\it ff}$}\label{appendix-GA2UGAFF}

In this Appendix, we prove Lemma \ref{GA2UGAFF}, in which UniqueGA$_{\it ff}$ is shown to be polynomial-time Turing equivalent to GA. 
Earlier, K{\" o}bler, 
Sch{\" o}ning, and Tor{\' a}n~\cite{KST93} established 
the polynomial-time Turing 
equivalence between 
GA and UniqueGA. We first review their reduction and then explain  
how to modify it to obtain the desired reduction from GA to UniqueGA$_{\it ff}$. Note that the reduction from UniqueGA$_{\it ff}$ to GA is trivial 
since UniqueGA$_{\it ff}$ is simply a special case of GA.
%since 
%a set of ``YES" instances of UniqueGA$_{\it ff}$ is contained in that of GA 
%and a set of ``NO" instances of UniqueGA$_{\it ff}$ is equal to that of GA.

We begin with explaining our technical tool and notation necessary to describe the reduction of \cite{KST93}. Their reduction uses a technical tool called a {\em label} to distinguish each node of a given graph $G$ from the others. 
Given a graph $G$, let $n$ be the number of nodes in $G$.
The label $j$ attached to node $i$ consists of two chains: 
one of which is of length 
$2n+3$ connected to node $i$, and the other is of length $j$ connected 
to the $n+2$-nd node of the first chain. (See Figure~2.)
%\vspace*{16mm}
%
%\begin{center}
%------------\\
%Figure~2\\
%------------
%\end{center}
%
%\vspace*{16mm}
%
\begin{center}
\resizebox{5cm}{!}{ 
\includegraphics*{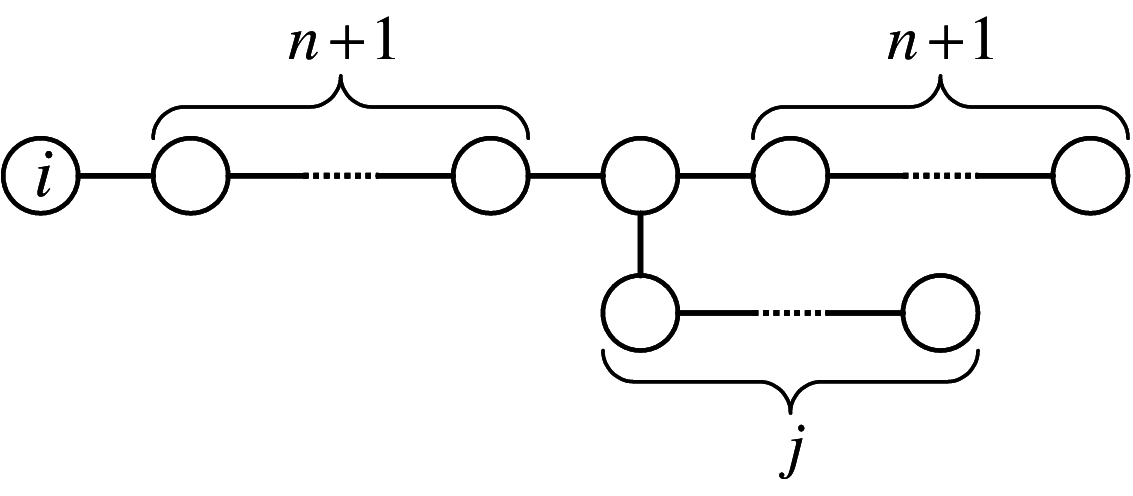}
}\\
Figure~2: label
\end{center}
Note that the total size of the label $j$ is $2n+j+3$. 
Let $G_{[i]}$ denote 
the graph obtained from $G$ by attaching the label $1$ to the node $i$. 
Similarly, $G_{[i_1,...,i_j]}$ is defined as the graph 
with labels $1,...,j$ respectively attached to nodes $i_1,...,i_j$.
Note that any automorphism of $G_{[i]}$ maps the node $i$ into 
itself and that any label adds no 
new automorphism into this modified graph.
Let $Aut(G)$ be the automorphism group of $G$ and let $Aut(G)_{[1,...,i]}$
be the point-wise stabilizer of $\{1,...,i\}$ in $Aut(G)$, namely, 
$Aut(G)_{[1,...,i]} = \{ \sigma\in Aut(G): \forall j\in\{1,...,i\}[\sigma(j) = j] \}$.

The following theorem was proven in \cite{KST93}. 
For our later reference, we include its proof here.

\begin{theorem}\label{GA-to-UniqueGA}{\rm \cite[Theorem 1.31]{KST93}}\ 
GA is polynomial-time Turing reducible to UniqueGA.
\end{theorem}

\begin{proof}
Let ${\cal O}$ be any set that correctly represents UniqueGA on all 
promised instances. Using ${\cal O}$ as an oracle, 
the following algorithm solves 
GA in polynomial time. Let $G$ be any given instance of GA.
\begin{description}
\magicwand
\item[(U1)] Repeat (U2)-(U3) for each $i$ starting with $n-1$ down to $1$.
\item[(U2)] Repeat (U3) for each $j$ ranging from $i+1$ to $n$.
\item[(U3)] Invoke ${\cal O}$ with input graph $G_{[1,...,i-1,i]} \cup G_{[1,...,i-1,j]}$. 
If the outcome of ${\cal O}$ is YES, output YES and halt.
\item[(U4)] Output NO.
\end{description}

If $G$ is an ``YES" instance, there is at least one non-trivial 
automorphism. Take the largest number $i\in\{1,...,n\}$ 
such that there exist a number $j\in\{1,...,n\}$ and a non-trivial automorphism $\pi\in Aut(G)_{[1,...,i-1]}$
for which $\pi(i)=j$ and $i\neq j$.
We want to claim that there is exactly one such 
non-trivial automorphism, i.e.,
$Aut(G)_{[1,...,i-1]}=\{id,\pi\}$.
This is seen as follows. First, note that $Aut(G)_{[1,...,i-1]}$ is expressed as 
$
 Aut(G)_{[1,...,i-1]} = \pi_1 Aut(G)_{[1,...,i]} + \cdots + \pi_d Aut(G)_{[1,...,i]}.
$
For any two distinct cosets $\pi_s Aut(G)_{[1,...,i]}$ and $\pi_t Aut(G)_{[1,...,i]}$ and for 
any two automorphisms $\sigma\in\pi_s Aut(G)_{[1,...,i]}$ and $\sigma'\in\pi_t Aut(G)_{[1,...,i]}$, it holds that $\sigma(i) \neq \sigma'(i)$.
Since $Aut(G)_{[1,...,i]}=\{id \}$ by the definition of $i$, we obtain
$|\pi_kAut(G)_{[1,...,i]}| = 1$ for any coset $\pi_kAut(G)_{[1,...,i]}$.
Furthermore, there exists the 
unique coset $\pi Aut(G)_{[1,...,i]}$ satisfying 
that $\sigma(i)=j$ for any 
$\sigma\in\pi Aut(G)_{[1,...,i]}$. 
These facts imply that the  
non-trivial automorphism $\pi$ is unique. 
Note that the unique non-trivial 
automorphism interchanges 
two subgraphs $G_{[1,...,i-1,i]}$ and $G_{[1,...,i-1,j]}$.
Therefore, the above algorithm successfully outputs YES at Step (U3).

On the contrary, if $G$ is a ``NO" instance, then for every distinct $i$ and $j$, 
the modified graph has no non-trivial automorphism. Thus, the above algorithm
correctly rejects $G$. 
\end{proof}

Finally, we describe the reduction from GA to UniqueGA$_{\it ff}$ 
by slightly modifying the reduction given in the above proof.

\begin{lemma}
GA is polynomial-time Turing reducible to UniqueGA$_{\it ff}$.
\end{lemma}
\begin{proof}
Recall the algorithm given in the proof of Theorem \ref{GA-to-UniqueGA}. 
We only need to change the number of nodes to 
invoke oracle UniqueGA$_{\it ff}$ in (U3). 
To make such a change, we first modify the size of each label.
Since the number $m$ of all nodes of $G_{[1,...,i-1,i]} \cup G_{[1,...,i-1,j]}$ 
is even, if there is no $k$ such that $m = 2(2k+1)$, then
we add one more node appropriately to the original labels. We then attach our modified labels of length $2n+i+4$ and $2n+j+4$ to the nodes $i$ and $j$, 
respectively. Obviously, this modified graph satisfies the promise of UniqueGA$_{\it ff}$. Our algorithm therefore works correctly 
for any instance of GA. 
\end{proof}
\newpage
\ignore{ %ignore
\begin{figure}[h]
\begin{center}
\scalebox{0.5}{\includegraphics{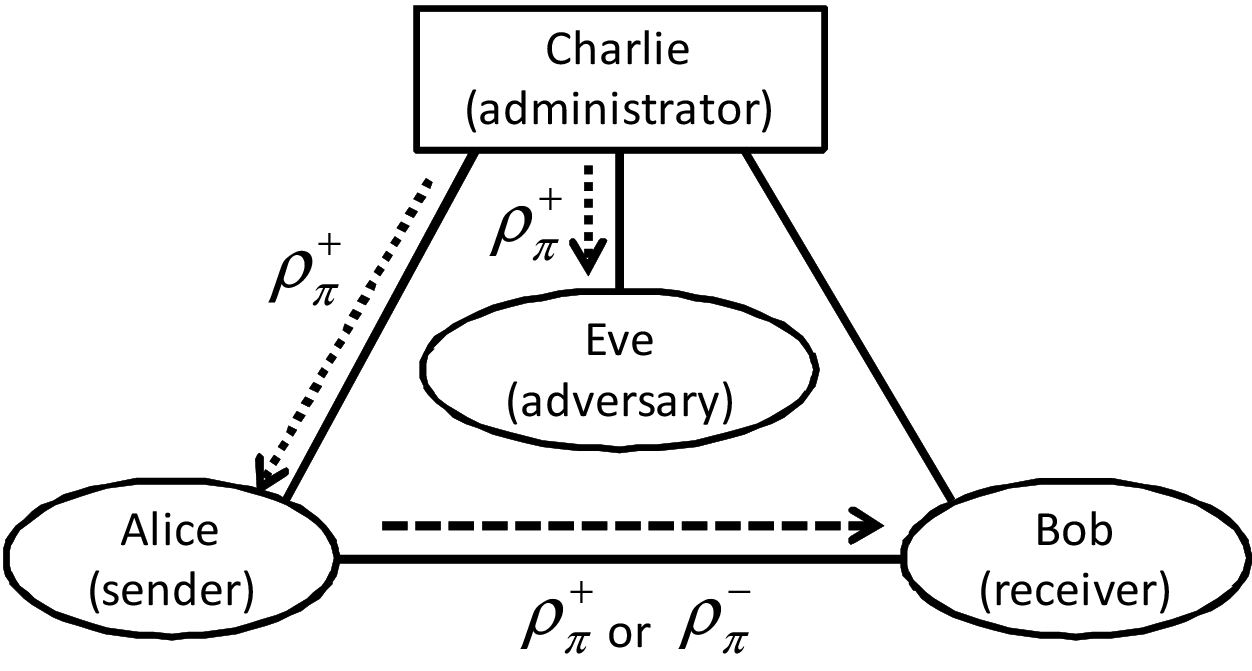}}\\
\caption{our public-key cryptosystem}
\end{center}
\end{figure}

\newpage

\begin{figure}[h]
\begin{center}
\scalebox{0.5}{\includegraphics{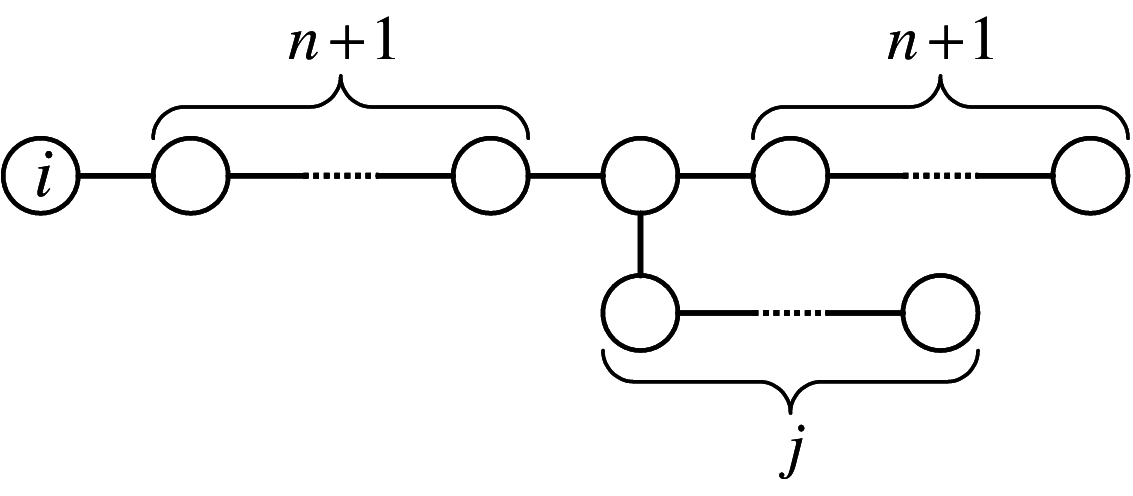}}\\
\caption{label}
\end{center}
\end{figure}
} %ignore

%%%%%%%%%%%%%%%%%%%%%%%%
\end{document}